# Two-to-three dimensional transition in neutral gold clusters: the crucial role of van der Waals interactions and temperature

Bryan R. Goldsmith,[1,2*] Jacob Florian,[2] Jin-Xun Liu,[2] Philipp Gruene,[1] Jonathan T. Lyon,[1,3] David M. Rayner,[4] André Fielicke,[1*] Matthias Scheffler,[1] and Luca M. Ghiringhelli[1*]

[1]*Fritz-Haber-Institut der Max-Planck-Gesellschaft, Faradayweg 4-6, D-14195 Berlin, Germany*
[2]*Department of Chemical Engineering, University of Michigan, Ann Arbor, MI 48109-2136, USA*
[3]*Department of Chemistry and Biochemistry, Kennesaw State University, 370 Paulding Avenue NW, MD#1203, Kennesaw, GA 30144, USA*
[4]*National Research Council, 100 Sussex Drive, Ottawa, Ontario, Canada K1A OR6*
Corresponding authors: bgoldsm@umich.edu, fielicke@fhi-berlin.mpg.de, ghiringhelli@fhi-berlin.mpg.de

We predict the structures of neutral gas-phase gold clusters (Au$_n$, $n$ = 5–13) at finite temperatures based on free-energy calculations obtained by replica-exchange *ab initio* molecular dynamics. The structures of neutral Au$_5$–Au$_{13}$ clusters are assigned at 100 K based on a comparison of experimental far-infrared multiple photon dissociation spectra performed on Kr-tagged gold clusters with theoretical anharmonic IR spectra and free-energy calculations. The critical gold cluster size where the most stable isomer changes from planar to nonplanar is Au$_{11}$ (capped-trigonal prism, D$_{3h}$) at 100 K. However, at 300 K (i.e., room temperature), planar and nonplanar isomers may coexist even for Au$_8$, Au$_9$, and Au$_{10}$ clusters. Density-functional theory exchange-correlation functionals within the generalized gradient or hybrid approximation must be corrected for long-range van der Waals interactions to accurately predict relative gold cluster isomer stabilities. Our work gives insight into the stable structures of gas-phase gold clusters by highlighting the impact of temperature, and therefore the importance of free-energy over total energy studies, and long-range van der Waals interactions on gold cluster stability.

Metal clusters in the gas phase are widely used as model systems to study fundamental properties of condensed matter (see, e.g., Refs. [1-7]). At the nanoscale, gold is not a fully noble, nonreactive material; thus gold clusters are of particular interest due to their possible applications in gas sensing, pollution reduction, and catalysis [8-17]. Gold clusters in the gas phase exhibit many structural isomers of similar energetics and can rapidly interconvert among them [18-24]. Geometry and size can impact the physicochemical properties of clusters, for example, HOMO-LUMO gap, polarizability, and catalytic activity [25-31]. Small gold clusters often adopt stable planar geometries [32-36]. The critical size where gold clusters begin favoring nonplanar (three-dimensional, 3D) structures over planar (two-dimensional, 2D) structures has attracted sustained interest [37-44]. Gas-phase ion-mobility experiments at room temperature suggest a transition from a 2D to 3D ground state structure at size 12 [45] and 8 [46] for negatively and positively charged clusters, respectively. Computational and experimental evidence supports these ground-state structural assignments for the 2D to 3D transition size [47,48]. For neutral gold clusters, computational studies predict their critical transition size is between Au$_{10}$ and Au$_{14}$ at zero kelvin [26,38,44], but experimental evidence is lacking due to the difficulty in spectroscopically characterizing neutral gas-phase clusters relative to charged clusters.

Both experimental and computational studies indicate that dynamic structural rearrangements are a common feature among clusters at finite temperatures [22,49-57]. It is clear that small gold clusters favor a 2D structure. The critical size where gold clusters favor 3D structures over 2D structures likely depends on temperature, but few studies have computationally examined the isomer stability at finite temperature [49,57,58]. One study used metadynamics to probe the free-energy surface of neutral Au$_{12}$ and predicted dynamic coexistence of multiple planar and nonplanar isomers at room temperature [58].

Besides temperature effects on cluster stability, the long-range tail contributions of van der Waals interactions (from here on we simply abbreviate as vdW) in Kohn-Sham density functional theory (DFT) studies of clusters have been neglected in most previous works. Van der Waals interactions can be crucial for predicting the stability of materials and molecules [59-63], molecule adsorption to surfaces [64], and even selectivity of reactions [65]. Including long-range vdW interactions in DFT calculations has been suggested to stabilize 3D isomers relative to 2D isomers for small gold clusters [57,66,67], nonetheless, the magnitude of long-range vdW interactions on gold cluster isomer stability over a broad size range has not been quantified. Ultimately, accurate first-principles predictions, including both temperature effects (i.e., enthalpic and entropic contributions to obtain free energies) and vdW interactions, combined with experimental characterization, are needed to unambiguously identify the structures of clusters at finite temperature.

The structures of Au$_3$–Au$_8$ at 100 K have been previously determined by some of the authors via a combination of both first-principles modeling and far-infrared multiple photon dissociation (FIR-MPD) spectroscopy [55,66]. Here, we predict the structures of neutral gas-phase gold clusters (Au$_n$, $n$ = 5–13) at finite temperatures based on free-energy calculations using replica-exchange *ab initio* molecular dynamics combined with the Multistate Bennett Acceptance Ratio (REMD+MBAR) [68]. We then compare experimental FIR-MPD spectra performed on Kr-tagged gold clusters with theoretical anharmonic IR spectra and free-energy calculations to structurally assign $Au_9$–$Au_{13}$, where the transition from planar to three-dimensional structures is expected to occur for neutral gold clusters.

## METHODS

*Computational:* The FHI-aims package is used for all electronic-structure calculations [69,70]. FHI-aims gives an accurate all-electron, full-potential description based on numeric atom-centered basis functions. Since the gold cluster ground-state structures are expected to depend on the theoretical method used [44,48], the lowest energy isomers are examined at various levels of DFT and beyond. The exchange-correlation (xc) density-functional approximations (referred to as DFA from here on) that we use are the Perdew-Burke-Ernzerhof (PBE) xc functional [71] and the Heyd-Scuseria-Ernzerhof (HSE06) hybrid xc functional [72,73]. Because these semi-local xc functionals do not account for the long-range vdW interactions, here we augment our DFAs with long-range vdW corrections using both the pairwise Tkatchenko-Scheffler (TS) [74] method and the many-body dispersion (MBD) approach with range separated self-consistent screening [75,76]. We call these approximations DFA+TS and DFA+MBD, where DFA is replaced by the name of the xc functional used to approximate the energy. For HSE06 calculations we use a screening parameter of 0.11/bohr and 25% exact exchange [77].

Geometry optimizations used (collinear) spin-polarized DFT with "tight" integration grid settings and accurate "tier 2" basis sets, unless stated otherwise. All DFA+TS and DFA+MBD geometry optimization calculations are compared against RPA (exact-exchange plus electronic correlation in the <u>r</u>andom-<u>p</u>hase <u>a</u>pproximation) single-point calculations, starting from PBE orbitals (denoted as RPA@PBE) [78]. RPA calculations have been used before to study the stability of gas phase gold clusters [44]. Our RPA@PBE calculations were corrected for basis-set superposition error and used "very-tight" integration grid settings with the large cc-pV5Z-PP basis from the EMSL basis-set exchange database [79]. Zero-point energy corrections are not included as they are typically not required for atomic clusters. Gold relativistic effects are treated using the atomic ZORA scalar correction [80]. Spin-orbit coupling is not included in any calculations [37,81]. A prior study suggested spin-orbit coupling only marginally affects the predicted ground-state structure for small gold clusters [37].

We use replica-exchange *ab initio* molecular dynamics (REMD) because it allows for an unbiased search of the potential energy surface from which gold cluster structures can be identified [82]. REMD simulations used energies and forces obtained from spin-polarized DFT using PBE+MBD with "light" integration grid settings and "tier 1" basis sets. REMD simulations used between 10 to 15 replicas distributed over the temperature range 100–1000 K. Each replica was initialized as a randomly generated structure. Between attempted replica exchanges, constant-temperature, constant-volume (i.e., NVT) Born-Oppenheimer *ab initio* molecular dynamics ran for 0.5 ps using a 10 fs time step. The simulation time per replica was 300 ps, resulting in a combined simulation time of at least 3 ns for each $Au_n$ cluster ($n$ = 5–13). The stochastic velocity rescaling thermostat was used to sample the canonical (NVT) ensemble with a $\tau$ parameter of 100 fs [83]. Cluster configurations and potential energies were subsampled (with lag time based on the decay of the velocity autocorrelation of the simulation trajectories) from the REMD simulations to obtain statistically independent samples to compute free-energy surfaces via the Multistate Bennett Acceptance Ratio (MBAR) using the pyMBAR software [68]. MBAR is a direct extension of BAR that allows for assessing data from all REMD states (here, a state is identified by the temperature) to predict free energies. The dimensionless free energies ($\beta\Delta F$, where $\beta = 1/k_BT$) for $Au_5$–$Au_{13}$ have an average error of $1.1\pm0.3$ $\beta\Delta F$. All the gold cluster configurations can be downloaded from the NOMAD Repository (http://dx.doi.org/10.17172/NOMAD/2016.11.02-1) [84].

Theoretical anharmonic IR spectra of $Au_9$–$Au_{13}$ at 100 K were computed from Born-Oppenheimer *ab initio* molecular-dynamics trajectories in the canonical ensemble (using the stochastic velocity rescaling thermostat) by evaluating the ensemble average of the Fourier transform of the dipole-dipole time-autocorrelation function [55,85,86]. In this way, anharmonic effects are fully included in the calculations. These simulations were 100–150 ps in duration and used energies and forces obtained from spin-polarized DFT with PBE+MBD and "tight" integration settings and "tier 2" basis sets. Although Kr is present in the FIR-MPD spectroscopy experiments, we neglect Kr in our theoretical anharmonic IR spectra because its impact on the gold cluster vibrational spectra is likely negligible at the cluster sizes considered in this study – the effect of Kr to the IR spectrum for $Au_7$ was minimal in a prior study by some of the authors [55]. Pendry reliability factors (Pendry R-Factor, $R_P$) were computed to give a quantitative comparison of how well theoretical IR spectra peak positions and their intensities agree with experimental FIR-MPD spectra. $R_P$ compares peak positions between two spectra using the renormalized logarithmic derivatives of their intensities [87], which results in $R_P = 0$ for perfectly agreeing spectra, $R_P = 1$ for uncorrelated spectra, and $R_P = 2$ for anticorrelated spectra. Because $R_P$ is sensitive to noise and needs well-separated peaks, we smoothed the experimental FIR-MPD spectra four times and theoretical IR spectra twice using a three-point formula with a 0.6 cm$^{-1}$ grid separation [88]. A constant half width max of 5 cm$^{-1}$ was used when computing $R_P$. A small frequency shift instead of a scaling factor was applied to the theoretical IR spectra, specifically: 8 cm$^{-1}$ ($Au_9$), 8 cm$^{-1}$ ($Au_{10}$), 12 cm$^{-1}$ ($Au_{11}$), 3 cm$^{-1}$ ($Au_{12}$), and 8 cm$^{-1}$ ($Au_{13}$) – see Ref. [55] and references within for discussion on this issue.

*Experimental:* Vibrational spectra are obtained by FIR-MPD spectroscopy of complexes of neutral gold clusters with Kr atoms. In short, the gas-phase clusters are produced via laser ablation and thermalized to ~100 K. By adding 1.5% Kr to the He carrier gas, formation of Kr complexes of the neutral gold clusters is achieved. By ionization with an $F_2$ laser (157 nm, 7.9 eV) the cluster distribution can be analyzed using time-of-flight mass spectrometry. Before reaching the mass spectrometer the neutral cluster beam is irradiated by light pulses from the Free Electron Laser for Infrared eXperiments (FELIX) [90]. If at a given IR frequency a cluster complex has an IR active mode, it can absorb photons, eventually leading to dissociation of the complex. This intensity change in the neutral cluster distribution is detected after ionization by the same relative intensity change in the mass spectrum. By scanning the frequency of FELIX, mass selective FIR-MPD spectra of the Kr-tagged gold clusters are obtained from the IR induced intensity change $I/I_0$ and plotted as experimental cross section $1/P \ln I_0/I$ with $P$ being the IR laser power at a given wavenumber. More details on the experimental methods have been given before [66].

## RESULTS AND DISCUSSION

### Impact of density functional approximations and van der Waals interactions on gold-cluster stability

Accurate first-principles predictions are needed to discern the stabilities of cluster isomers at finite temperatures. Thus, we first briefly discuss the impact of selecting various DFAs for predicting the relative energetics of low-energy planar and nonplanar structures for neutral $Au_5$–$Au_{13}$ clusters. We compare our DFA calculations relative to RPA. RPA is an attractive method for comparison because the RPA correlation energy is fully non-local and includes long-range vdW interactions and dynamic electronic screening, and the exact-exchange energy fully cancels the self-interaction error present in the Hartree energy [78]. Ultimately, the planar and nonplanar $Au_5$–$Au_{13}$ structures predicted to be most stable at zero kelvin are shown in **FIG. 1**.

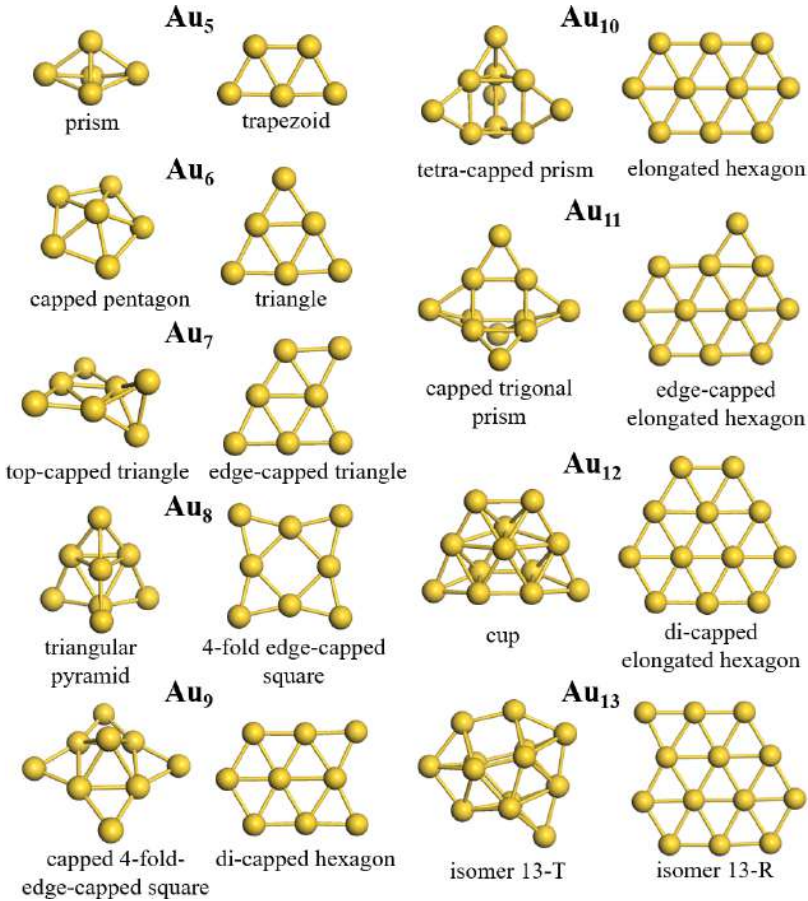

**FIG. 1.** Most stable planar and nonplanar gold cluster structures for $Au_5$–$Au_{13}$ at zero kelvin. The same low-energy structures are identified for all considered density functional approximations (e.g., using either PBE+MBD or HSE06+MBD) and RPA.

Energy differences between the most stable nonplanar and planar isomers of $Au_5$–$Au_{13}$ (structures corresponding to **FIG. 1**) using various DFAs and RPA@PBE are shown in **FIG. 2**. Geometries were relaxed for each cluster with the respective DFA, but RPA@PBE was used as a single point correction on top of PBE-optimized geometries. Energetics corresponding to **FIG. 2** are provided within **Table S1** in the supporting information (SI). The preference toward favoring 3D gold isomers over 2D isomers generally increases as cluster size increases. For $n$ = 5–10 and 12, all DFAs predict planar structures are the most stable, but for $n$ = 11 and 13, there are differences. The critical transition size from 2D to 3D at zero kelvin is above $Au_{13}$ using PBE, whereas for all other DFAs and RPA@PBE it is $Au_{11}$. The 2D to 3D critical transition size at zero kelvin was previously predicted at $Au_{11}$, where $Au_{11}$ preferred a capped-trigonal prism ($D_{3h}$) structure [44,89], which is the same nonplanar isomer we found (**FIG. 1**). Although $Au_{11}$ prefers a nonplanar isomer at zero kelvin, all our DFA and RPA@PBE calculations predict $Au_{12}$ to prefer a planar di-capped elongated hexagon ($D_{3h}$). Another first-principles study also found $Au_{12}$ prefers a planar di-capped elongated hexagon at zero kelvin [67], whereas other studies predict a nonplanar "cup" ($C_{2v}$) structure as the preferred isomer [44]. As we discuss further below, the comparison of the experimental FIR-MPD spectra to theoretical IR spectra identifies the $Au_{12}$ nonplanar cup isomer at 100 K. PBE+MBD, HSE06+MBD, and RPA@PBE predict the same ground-state structure for $Au_5$–$Au_{12}$, but PBE+MBD slightly favors a planar $Au_{13}$ isomer ($\Delta E_{3D \to 2D}$ = −34 meV), whereas HSE06+MBD and RPA@PBE calculations favor a nonplanar $Au_{13}$ structure by 92 meV and 89 meV, respectively. Indeed, $Au_{13}$ has been suggested to be near the critical 2D to 3D transition size on multiple occasions, with much debate on the ground-state structure due to the sensitivity to the first-principles modeling approach [26,27,37,44,45].

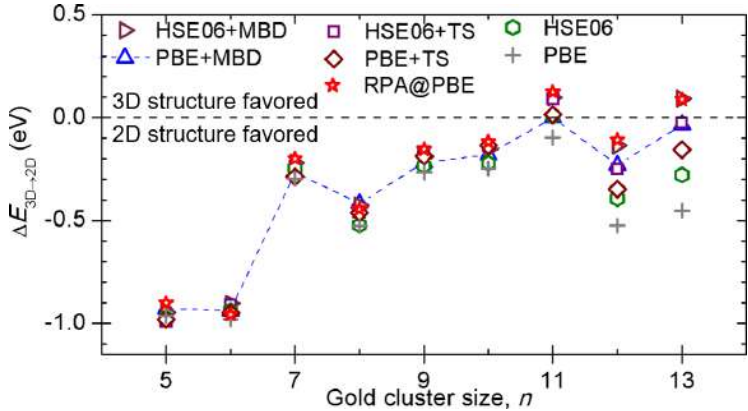

**FIG. 2.** Energy difference between the lowest energy nonplanar and planar isomers ($\Delta E_{3D \to 2D}$) for $Au_5$–$Au_{13}$ using various density functional approximations and beyond at zero kelvin. Dashed blue line is used to guide the eye for the PBE+MBD results. Abbreviations: TS = Tkatchenko-Scheffler pairwise van der Waals correction; MBD = Many-body dispersion with range separated self-consistent screening. RPA@PBE = random-phase approximation using PBE orbitals. 3D = nonplanar; 2D = planar.

For the data in **FIG. 2**, the DFA error with respect to RPA@PBE increases monotonically with gold cluster size. HSE06+TS and HSE06+MBD agree best with RPA@PBE among our tested DFAs (mean absolute error of 57±45 meV and 30±14 meV, respectively), but the computational cost of a hybrid functional prohibits their use in replica-exchange *ab initio* molecular dynamics simulations. Nevertheless, the results of **FIG. 2** show that HSE calculations are not needed for this study. The good tradeoff between accuracy and computational cost of PBE+MBD justifies its use in our REMD calculations to compute the free-energies of the structures of neutral $Au_5$–$Au_{13}$ clusters.

Interestingly, for $Au_5$–$Au_9$ all DFAs give essentially the same $\Delta E_{3D \to 2D}$, thus the non-vdW tail interactions favor the 2D structures. However, including long-range vdW interactions is important for larger clusters to improve agreement with RPA@PBE. HSE06 typically performs worse than PBE+TS and PBE+MBD with respect to RPA@PBE for predicting $\Delta E_{3D \to 2D}$ of $Au_5$–$Au_{13}$, having a MAE of 130±110 meV. Correcting PBE with MBD has a more pronounced effect on the relative isomer energetics than changing from PBE to HSE06. Going from PBE to PBE+TS or PBE+MBD decreases the MAE with respect to RPA@PBE by 94 meV and 116 meV, respectively. Similarly, going from HSE6 to HSE06+TS or HSE06+MBD decreases the MAE with respect to RPA@PBE by 73 meV and 100 meV. The DFA+MBD scheme on average performs better than DFA+TS. This result is expected because the TS method simply computes the vdW energy using the ground-state electron density and includes hybridization effects for the polarizability but neglects long-range electrostatic screening and many-body contributions beyond pairwise interactions. However, the MBD scheme corrects the issues of TS by including the screened long-range many-body vdW energy via solution of the Schrödinger equation for a system of coupled electronic oscillators [76].

None of the vdW corrections applied in this study is expected to yield a full account of vdW interactions among Au atoms [90], due to delocalized 6s electrons, which challenge the charge partitioning scheme behind the adopted vdW-correction methods. However, we note the remarkable quantitative agreement between the vdW-corrected DFAs, and in particular HSE06+MBD and the RPA when the relative energy between 2D and 3D structures are compared (**FIG. 2**). This gives us confidence that the essence of the relative stability between planar and nonplanar $Au_n$ structures is captured by the adopted vdW-correction schemes, in particular MBD.

Our results show that including long-range vdW interactions stabilizes 3D structures relative to 2D structures, especially as cluster size increases above seven atoms, **FIG. 3**. For example, using PBE+MBD the vdW interactions stabilize 3D structures relative to 2D structures by 110 meV for $Au_8$ and 456 meV for $Au_{13}$. Similar trends are found using HSE06+MBD and the TS scheme with and without self-consistent screening (SCS) to account for electrodynamic response effects.

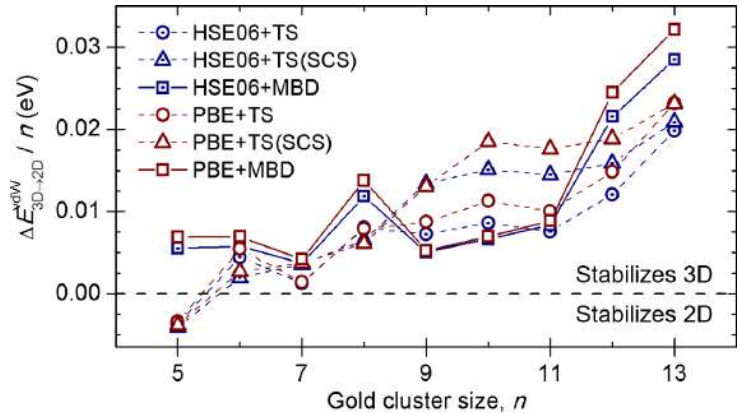

**FIG. 3.** Difference in van der Waals energy per atom ($\Delta E^{vdW}_{3D \rightarrow 2D} / n$) between the most stable planar and nonplanar isomers of $Au_5$–$Au_{13}$ at zero kelvin using various density-functional approximations and van der Waals corrections. Abbreviations: TS = Tkatchenko-Scheffler pairwise van der Waals; TS(SCS) = Tkatchenko-Scheffler pairwise van der Waals with self-consistent screening; MBD = Many-body dispersion with range separated self-consistent screening; 3D = nonplanar; 2D = planar.

We next predicted the average isotropic static polarizability per atom ($\alpha_{iso}$) of the lowest energy planar and nonplanar gold clusters as a function of size using HSE06+MBD(SCS) and HSE06+TS (**FIG. 4**). MBD(SCS) properly captures the static polarizability, whereas the TS scheme does not. There is an odd-even oscillation with respect to cluster size using MBD(SCS), where odd-sized gold clusters typically show larger polarizabilities than their even-sized neighbors as expected for open/closed shell systems. Also, we predict that 2D clusters have a larger $\alpha_{iso}$ than 3D clusters. Nevertheless, even though the 2D clusters have a higher average $\alpha_{iso}$, the atoms of the 3D clusters have higher atomic coordination (due to their more compact nature), leading to stronger stabilization by vdW forces (**FIG. S1**. and **FIG. S2** of the SI).

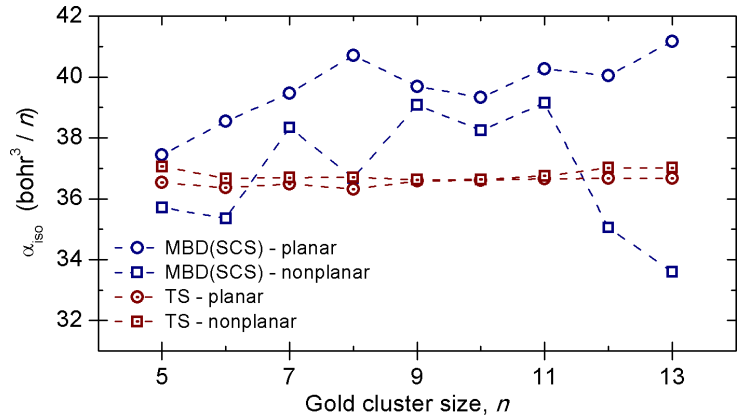

**FIG. 4.** Average isotropic static polarizability ($\alpha_{iso}$) per atom of the lowest energy planar and nonplanar gold clusters as a function of size predicted using HSE06+MBD(SCS) and HSE06+TS. The planar and nonplanar gold clusters correspond to those shown in FIG. 1. Abbreviations: TS = Tkatchenko-Scheffler pairwise van der Waals; MBD(SCS) = Many-body dispersion with self-consistent screening.

**Free-energy surfaces of $Au_5$–$Au_{13}$: impact of temperature on cluster isomer stability**

The usual modeling approach to probe the stability of cluster isomers is to compare the energies of various isomers at zero kelvin. However, experiments are performed at finite temperatures, thus structures with low free energy (high Boltzmann probability) should be determined. Calculating free energy of clusters, which include anharmonic contributions and the effect of frequent structural rearrangements (fluxionality), requires sampling many configurations to capture the enthalpic and entropic contributions to the free energy. Our REMD+MBAR simulations efficiently sampled the potential energy surfaces of neutral $Au_n$ clusters ($n$ = 5–13) without requiring knowledge *a priori* and let us compute free-energies of gold cluster isomers, and hence their Boltzmann probability. Gold cluster isomers and their Boltzmann populations are depicted in two-dimensional free-energy surfaces using two order parameters that were selected *a posteriori*, namely, the gold cluster's "coordination similarity" and "radius of gyration". Together, these two order parameters allow one to discern the different planar and nonplanar isomers in the simulation. The coordination similarity order parameter is obtained by constructing a coordination histogram (distribution of Au-atoms coordination number, i.e., number of Au atoms bonded to each Au atom) for each cluster configuration [91] and computing the cosine distance between each coordination histogram with respect to the ground-state reference (each cluster configuration referred to the ground state at the same size). Example bond coordination histograms for $Au_8$ at 100 and 300 K are provided in **FIG. S2** of the SI. Radius of gyration is computed by taking the root mean square distance of the cluster's atoms with respect to its center of mass.

**$Au_5$–$Au_8$.** Our free-energy favored structures fully agree with previous assignments for $Au_5$, $Au_6$, $Au_7$, and $Au_8$ based on FIR-MPD spectroscopy experiments (at 100 K) and ground-state DFT calculations [66,92]. $Au_5$ is most stable as a trapezoid ($C_{2v}$), having a Boltzmann probability ($P(T)$) of $P$(100 K) = 100% and $P$(300 K) = 99.2% (**FIG. S3** of the SI). The only thermodynamically competing structure at 300 K is a bow-tie ($D_{2h}$). $Au_6$ prefers a planar triangle ($D_{3h}$) structure with $P$(100 K) = 100% and $P$(300 K) = 100% according to the free-energy calculations (**FIG. S4** of the SI). The $Au_7$ cluster adopts the well-known edge-capped triangle ($C_s$) with $P$(100 K) = 100% and $P$(300 K) = 99.3%, with the second most likely structure being a hexagon ($D_{6h}$) with $P$(300 K) = 0.2% (**FIG. S5** of the SI). The remaining 0.5% of the $Au_7$ population is an assortment of other structures, indicative of the fluxional behavior of this system at 300 K. Nevertheless, for $Au_5$–$Au_7$ the impact of temperature on the isomer population is quite negligible up to 300 K, justifying the typical zero-kelvin modeling approach; however, for $Au_8$ the impact of temperature at 300 K is already quite noticeable. In agreement with past FIR-MPD experiments at 100 K [66] and computational predictions at zero kelvin [43], our free-energy calculations predict $Au_8$ prefers a planar 4-fold edge-capped square ($D_{4h}$) structure with $P$(100 K) = 100% (**FIG. S6** of the SI). Yet, at 300 K there is a competing nonplanar $Au_8$ isomer (triangular pyramid, $C_{3v}$) with $P$(300 K) = 5.4%. Thus, even for clusters as small as $Au_8$ there can be a thermodynamically competing 3D isomer with an appreciable Boltzmann probability at room temperature. As we show below, the impact of raising temperature above 100 K typically stabilizes nonplanar gold cluster structures over planar structures, with the exception of $Au_{11}$.

The perfect agreement between the here predicted structures of $Au_5$–$Au_8$ at 100 K and previous structural assignments from experimental FIR-MPD spectra at 100 K gives us confidence in our REMD+MBAR modeling approach using PBE+MBD. In the remainder of this section we analyze the free-energy surfaces of $Au_9$–$Au_{13}$ to examine isomer stability at finite temperatures. In the next section we compare the free-energy favored structures of $Au_9$–$Au_{13}$ with experimentally measured and theoretically predicted IR spectra to finally assign the structures of $Au_9$–$Au_{13}$ at 100 K.

**$Au_9$.** The free-energy surface of $Au_9$ shown in **FIG. 5** predicts the most stable isomer at both 100 K and 300 K is a di-capped hexagon with $C_{2v}$ symmetry (isomer **9-A**) with $P_{9-A}$(100 K) = 100% and $P_{9-A}$(300 K) = 92.4%. Previous first-principles calculations also predicted **9-A** to be the ground-state structure at zero kelvin [44]. **9-A** is clearly identified as the preferred $Au_9$ structure at 100 K according to our FIR-MPD spectroscopy measurements and theoretical IR spectra (discussed further below, **FIG. 10**). At 300 K, however, there are two competing isomers **9-B** and **9-C** with non-negligible Boltzmann probabilities of $P_{9-B}$(300 K) = 3.3% and $P_{9-C}$(300 K) = 2.5%. The nonplanar **9-B** is a capped 4-fold edge-capped square (built off the most stable $Au_8$ structure), whereas **9-C** is a planar trapezoid. **9-A** and **9-C**, having simply a different position of one periphery (edge) gold atom, have a more similar radius of gyration and coordination similarity compared with **9-B**. Although nonplanar isomers are thermodynamically stable for $Au_9$ even at 300 K, the barriers between the planar (**9-A** and **9-C**) and nonplanar (**9-B**) isomers are calculated to be much greater than 25 $\beta\Delta F$ (0.65 eV), thus $Au_9$ might be kinetically trapped in planar or nonplanar structures.

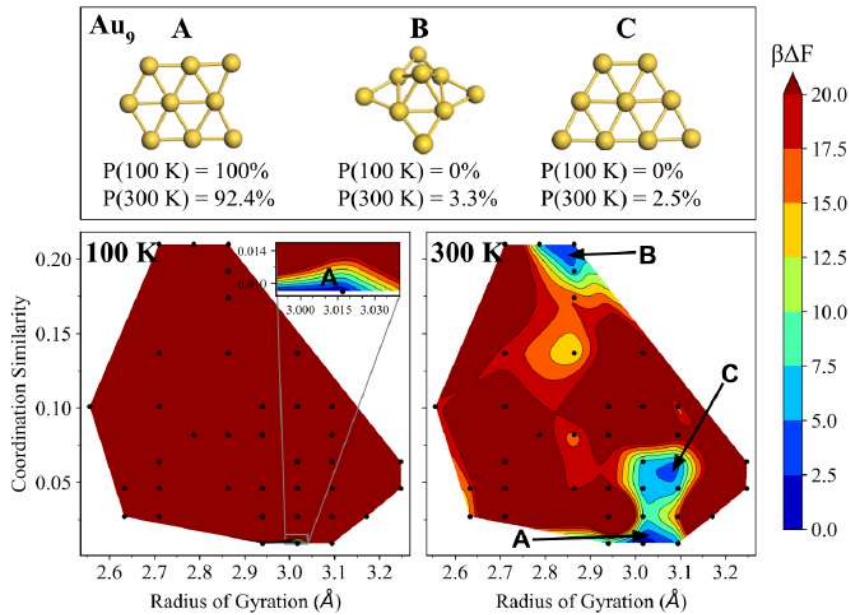

**FIG. 5.** Free-energy surfaces of Au$_9$ at 100 K (left) and 300 K (right). The dominant isomer **9-A** (di-capped hexagon, C$_{2v}$) and competing isomers **9-B** (capped 4-fold-edge-capped square, C$_{4v}$) and **9-C** (trapezoid, C$_{2v}$) have their Boltzmann probabilities $P(T)$ specified at each temperature, as well as their locations marked on the free-energy surfaces. The color bar represents dimensionless free-energy values ($\beta\Delta F$), where $\beta$ is $1/k_B T$. Dark red regions have $\beta\Delta F \geq 20$. The black points indicate sampled cluster geometries having the binned (coordination similarity, radius of gyration) value. **A** is used as the reference state for coordination similarity, thus its coordination similarity is 0.0.

**Au$_{10}$.** The free-energy surface of Au$_{10}$ shown in **FIG. 6** predicts the most stable isomer at 100 K is an elongated hexagon with D$_{2h}$ symmetry (**10-D**) with $P_{10\text{-D}}(100\text{ K}) = 99.3\%$. Previous first-principles calculations at zero kelvin also predicted **10-D** as the ground-state structure [44]. Moreover, **10-D** is identified as the preferred Au$_{10}$ structure at 100 K according to the assignment of the experimental FIR-MPD spectra (discussed further below, **FIG. 10**). However, there are multiple competing planar (**10-E**, **10-F**) and nonplanar (**10-G+**) isomers at 300 K. Interestingly, the most stable isomer **10-D** at 100 K is not the most stable isomer at 300 K, which is instead predicted to be **10-E**. Judging from the width and shallow nature of the free-energy basin of **10-E**, the stability of **10-E** over **10-D** at elevated temperature may arise from a larger configurational and vibrational entropy contribution to the free energy. Additionally, there are also three competing nonplanar isomers (**10-G+**) at 300 K with a combined probability of $P_{10\text{-G+}}(300\text{ K}) = 23.4\%$ (**G+** structures shown in **FIG. S7** of the SI). These three structures are not readily distinguishable by our order parameters, all having similar radius of gyration and coordination similarity, thus they are lumped into a single state. The broad and shallow free-energy basin of the nonplanar isomers **10-G+** suggests these structures can readily interconvert at room temperature, whereas the free-energy barrier to interconvert between planar and nonplanar isomers is likely much higher. At elevated temperatures of 500 K and 750 K, the various Au$_{10}$ planar and nonplanar isomers dynamically coexist as kinetic barriers can be readily overcome (**FIG. S8** of the SI).

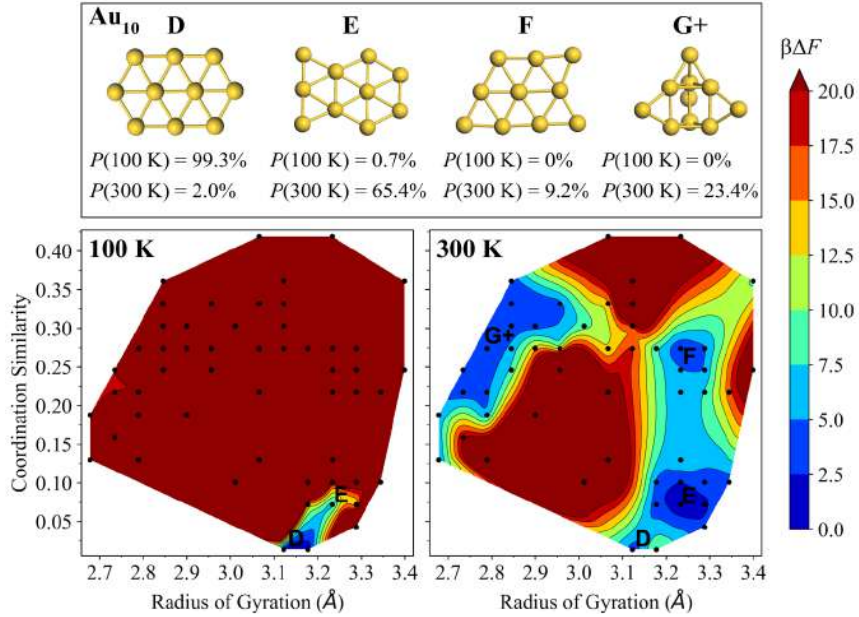

**FIG. 6.** Free-energy surfaces of Au$_{10}$ at 100 K (left) and 300 K (right). The dominant cluster isomers **10-D** (elongated hexagon, D$_{2h}$), **10-E** (tri-capped hexagon, C$_{2h}$), **10-F** (capped trapezoid, C$_s$), and **10-G+** have their Boltzmann probabilities $P(T)$ specified at each temperature, as well as their locations marked on the free-energy surface.

**Au$_{11}$.** For Au$_{11}$, several isomers, planar **11-H+** (differently edge-capped elongated hexagons) as well as nonplanar, **11-K** (capped trigonal prism) and **11-L**, have non-negligible Boltzmann probabilities at 100 K, **FIG. 7**. At 300 K, a variety of planar and nonplanar Au$_{11}$ structures can coexist, with roughly $P(300 \text{ K}) = 55\%$ for observing a planar structure. The metastable isomers (**11-I+**, **11-M**, **11-N+**) not shown in **FIG. 7** are shown in **FIG. S9** of the SI. The ensemble of planar structures of Au$_{11}$ resemble the planar structures of Au$_{10}$, except with an 11$^{th}$ atom that can easily migrate around the cluster periphery to adopt a variety of distinct and stable planar isomers. The nonplanar **11-K** has the lowest free energy at 100 K and is identified according to our experimental FIR-MPD and theoretical IR spectra (discussed below, **FIG. 10**), indicating that Au$_{11}$ is the critical 2D/3D transition size at 100 K.

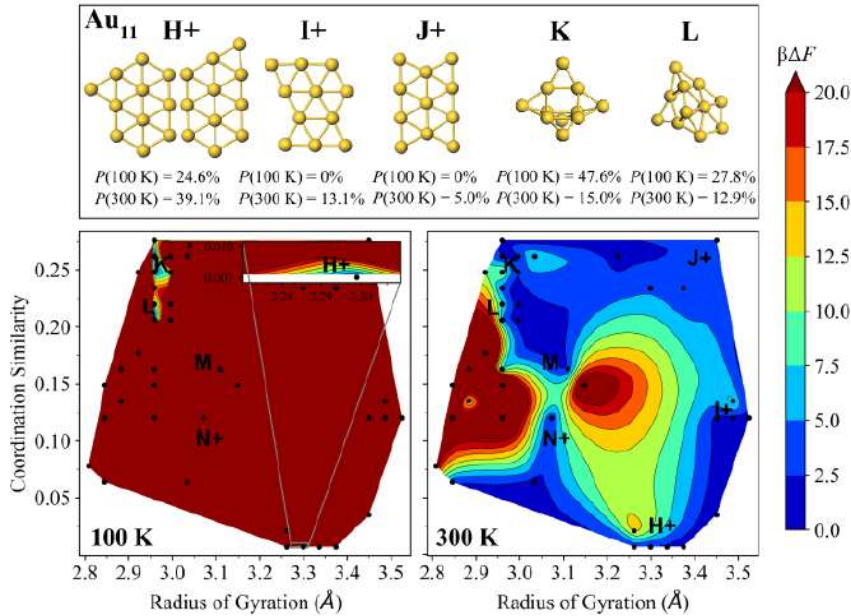

**FIG. 7.** Free-energy surfaces of Au$_{11}$ at 100 K (left) and 300 K (right). The dominant cluster isomers **11-H+**, **11-I+**, **11-J+**, **11-K**, and **11-L** have their Boltzmann probabilities $P(T)$ specified at each temperature, as well as their locations marked on the free-energy surface. Isomers with $P(300 \text{ K}) < 1\%$ are not shown.

**Au$_{12}$.** Santarossa *et al.* analyzed the free-energy surface of Au$_{12}$ via metadynamics using PBE+Grimme-D1 and found a nearly equal probability for 2D and 3D isomers at room temperature, whereas the lowest energy 2D isomer is 0.2 eV more stable than the lowest energy 3D structure at zero kelvin [58]. In contrast, our free-energy calculations predict the nonplanar isomers will dominate at room temperature, **FIG. 8**. The most stable planar (**12-O**) and nonplanar (**12-P**) isomers we predict match those of Refs. [2,44,67] at zero kelvin; however, Santarossa *et al.* predicted **12-O** and **12-P** to be 0.65 eV and 1.64 eV uphill from the global minima, respectively. Although at 100 K the planar di-capped elongated hexagon (**12-O**) is most stable with $P_{12\text{-O}}(100\text{ K}) = 99.9\%$, at 300 K the nonplanar **12-P** and **12-Q** have $P_{12\text{-P}}(300\text{ K}) = 71.1\%$ and $P_{12\text{-Q}}(300\text{ K}) = 27.5\%$, respectively.

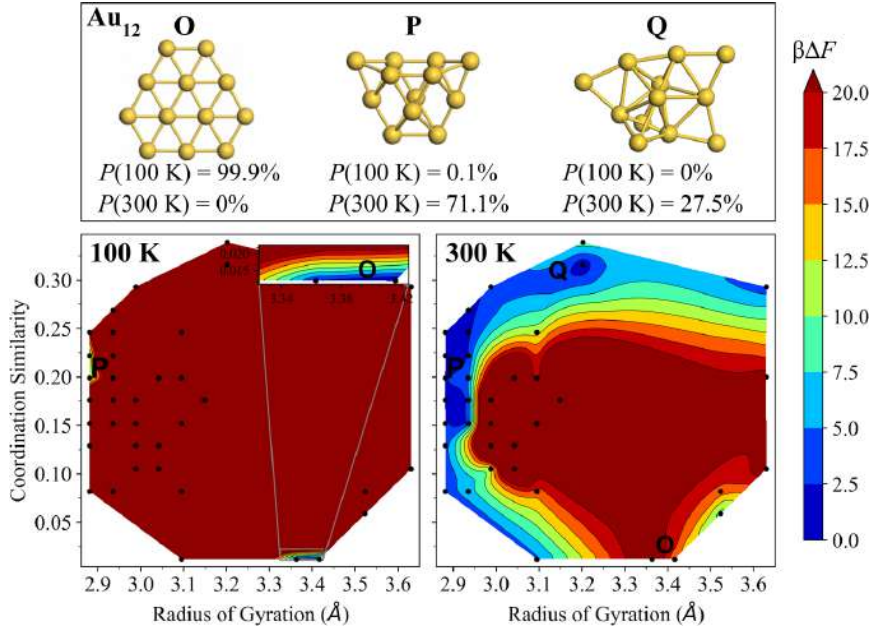

**FIG. 8.** Free-energy surfaces of Au$_{12}$ at 100 K (left) and 300 K (right). The dominant cluster isomers **12-O** (di-capped elongated hexagon, D$_{3h}$), **12-P** (cup, C$_{2v}$), and **12-Q** (C$_s$) have their Boltzmann probabilities $P(T)$ specified at each temperature, as well as their locations marked on the free-energy surfaces. Isomers with $P(300\text{ K}) < 1\%$ are not shown.

**Au$_{13}$.** Using PBE+MBD, we find that Au$_{13}$ at 100 K prefers a planar structure with $P(100\text{ K}) = 86.2\%$, although a variety of nonplanar structures has appreciable probabilities (**FIG. S10** of the SI). Although a few planar Au$_{13}$ isomers have a similar total energy to that of the lowest energy nonplanar Au$_{13}$ isomers, they were rarely seen due to temperature effects at 300 K. At 300 K a variety of nonplanar isomers are preferred with a cumulative probability of $P(300\text{ K}) = 90.5\%$. Indeed, low-energy isomers of Au$_{13}$ are known to be disordered, forming a nearly continuous distribution of nonplanar structures as a function of potential energy [50,57,93].

The free-energy surface can be calculated at any temperature within our REMD temperature-simulation window using MBAR. Based on integration of the free-energy surfaces of Au$_5$–Au$_{13}$, we show the impact of raising temperature from 100 K to 400 K on the Boltzmann population of planar structures vs. nonplanar structures, **FIG. 9**. Small gold clusters of neutral Au$_5$–Au$_7$ remain predominantly planar even as temperature is raised from 100 K at 400 K. However, increasing temperature from 100 K to 400 K typically stabilizes 3D structures over 2D isomers for Au$_8$–Au$_{13}$, with the exception of Au$_{11}$ due to its large number of nearly energy degenerate 2D structures that become kinetically accessible at elevated temperatures. Small populations of 3D isomers arise for Au$_8$ and Au$_9$ at 300 K and above (*ca.* 5% and 10% at 300 K and 400 K, respectively).

For Au$_{10}$-Au$_{13}$, we identified the temperature $T_{\text{2D/3D}}$ where there are roughly 50% planar and 50% nonplanar isomers. For Au$_{10}$ at $T_{\text{2D/3D}} = 750$ K, we compute a 48%±13% population of planar structures (error bars are based on uncertainty propagation of the free energies and neglecting correlations across free-energy bins). Although Au$_{11}$ slightly prefers a nonplanar structure over a planar structure at 100 K, a 52%±13% population of planar structures is calculated at $T_{\text{2D/3D}} = 115$ K. Au$_{12}$ and Au$_{13}$ adopt a 51%±12% and 50%±6.3% population of planar structures at 138 K and 155 K, respectively. The temperature where Au$_5$-Au$_9$ adopt roughly 50% planar structures is outside of our simulated temperature range (e.g., Au$_8$ and Au$_9$ only reach *ca.* 65% planar structures at 1000 K) and may not be achievable due to fragmentation at more elevated temperatures, i.e., there is effectively no planar/nonplanar transition for Au$_5$-Au$_9$. These results highlight the critical transition size from planar to nonplanar is gradual and depends on temperature.

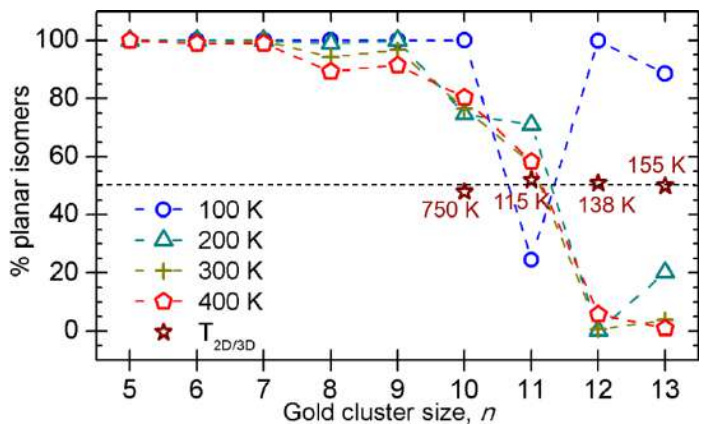

**FIG. 9.** Boltzmann population of planar isomers for $Au_5$–$Au_{13}$ as a function of temperature (100 K – 400 K) based on free-energy calculations. $T_{2D/3D}$ corresponds to the temperature where there are roughly 50% planar and 50% nonplanar structures.

**Structural assignment of $Au_9$–$Au_{13}$ via far-IR spectroscopy**

Most prior experimental studies on the structure of gas-phase gold clusters examined charged clusters due to the ease of size selection. Here we assign the structure of neutral $Au_9$–$Au_{13}$ clusters using experimental FIR-MPD spectroscopy and comparison to theoretical anharmonic IR spectra. FIR-MPD spectroscopy is a verified experimental technique for obtaining the vibrational spectra of metal clusters in the gas phase [94,95]. Some of the authors have previously structurally assigned the neutral $Au_3$, $Au_4$, $Au_5$, $Au_7$, $Au_8$, $Au_{19}$, and $Au_{20}$ clusters using FIR-MPD spectroscopy [55,66,92].

In most cases, the experimental FIR-MPD spectra of metal clusters were interpreted only by comparison to calculated harmonic IR spectra at zero kelvin. Experiments, however, are performed at finite temperature and even at low temperatures anharmonic effects can impact the vibrational spectra by changing the position of peaks, their intensities and broadening, and even cause the appearance of new peaks. Here, theoretical IR anharmonic spectra (from here on simply called theoretical IR spectra) of $Au_9$–$Au_{13}$ at 100 K are computed using *ab initio* molecular dynamics, which properly includes vibrational anharmonicity, to compare with experimental FIR-MPD spectra obtained at 100 K. **FIG. 10** and **FIG. 11** show the experimental FIR-MPD and theoretical IR spectra for $Au_9$–$Au_{13}$. Here we briefly note possible sources of discrepancy between the experimental FIR-MPD spectra and the theoretical IR spectra. First, the experimental FIR-MPD spectra may only be sampling the colder part of the canonically distributed population of neutral gold clusters due to spontaneous dissociation of the $Au_nKr_m$ complexes from the hotter tail of the canonical distribution [55]. Second, it is possible for fragmentation of $Au_nKr_m$ into $Au_nKr_{m-1}$ complexes to occur, which would affect the band intensities for $Au_nKr_{m-1}$ [96]. Also, only the pure metal clusters are considered in the calculations as it can be assumed that the weak interaction between Kr and the larger Au clusters does not has a significant influence on the IR spectra [55].

Johansson *et al.* predicted using DFT that the planar $Au_9$ di-capped hexagon, the planar $Au_{10}$ elongated hexagon, and the nonplanar $Au_{11}$ capped-trigonal prism are the ground-state structures at zero kelvin [44]. These same $Au_9$–$Au_{11}$ structures are assigned at 100 K based on comparison of our free-energy calculations and theoretical IR spectra with experimental FIR-MPD spectroscopy (**FIG. 10**).

**$Au_9$.** The $Au_9$ di-capped hexagon (**9-A**) is predicted to be the most probable at 100 K ($P_{9-A}(100\ K) = 100\%$) and also has the best theoretical IR spectrum agreement with FIR-MPD measurements for $Au_9Kr_2$ (**FIG. 10**) according to the Pendry R-Factor ($R_P = 0.80$). All other competing $Au_9$ isomers have 0% Boltzmann probabilities at 100 K and worse theoretical IR spectra agreement based on Pendry analysis. Albeit lacking the fine structure of the theoretical anharmonic IR spectra, the theoretical harmonic IR spectra of $Au_9$ (**FIG. S11** of the SI) also supports this structural assignment. Note that the experimental spectrum of $Au_9Kr$ shows in comparison to $Au_9Kr_2$ some broader less resolved bands around 180 cm$^{-1}$. This is probably due to a fragmentation of $Au_9Kr_2$ into $Au_9Kr$ affecting the band intensities for $Au_9Kr$. Alternatively one could assume that in $Au_9Kr$ a different structural isomer prevails, i.e., the rare gas binding may be isomer selective as seen for other gold clusters before [96]. However, comparison to the theoretical IR spectra of other isomers does not give indications for the presence of a different isomer.

**$Au_{10}$.** For $Au_{10}$, there is no significant difference between the FIR-MPD spectra of the complexes with a single and two Kr atoms and Pendry analysis between the FIR-MPD spectra and theoretical IR spectra indicates that either a nonplanar structure (**10-G**, $R_P = 0.80$) or the planar elongated-hexagon is preferred (**10-D**, $R_P = 0.81$). The theoretical IR spectrum of the **10-G** nonplanar isomer, however, is missing prominent peaks seen in the experiment at ~90 cm$^{-1}$ and ~160 cm$^{-1}$, which are present, however, in the theoretical IR spectrum of **10-D** and the FIR-MPD spectra of $Au_{10}Kr$ and $Au_{10}Kr_2$. This highlights a weakness of the Pendry analysis, which heavily weights peak positions over peak intensities. Thus, it is also important to also consider the free energy of the complexes beyond $R_P$ to aid structural assignments. The nonplanar **10-G** structure is predicted to have a 0% Boltzmann probability at 100 K. The planar **10-D** has the lowest calculated free energy at 100 K and has the second best agreement with the FIR-MPD spectra according to $R_P$. Thus, for $Au_{10}$ we assign the FIR-MPD spectra to the planar elongated-hexagon (**10-D**). Based on this result, one may assign $Au_{10}$ as the largest neutral gold cluster having a planar structure at 100 K, however, one has to keep in mind that the experiment is based on the investigation of the Kr complexes and, at least for anionic Au clusters, the planar isomers are known to favor complex formation with rare gas atoms. This has been used before to selectively measure the photoelectron spectrum of a minority 2D isomer of $Au_{12}^-$ out of a mixture of 2D/3D isomers [96]. Nevertheless, the experimental data is in excellent agreement with the theoretical IR spectra of **10-D** predicted to be most stable by the free-energy calculations.

**$Au_{11}$.** The nonplanar capped-trigonal prism (**11-K**, $R_P = 0.80$) of $Au_{11}$ shows excellent agreement of the theoretical IR spectrum with the experimental spectrum of $Au_{11}Kr_2$ having both two prominent peaks around 190 cm$^{-1}$ and 110 cm$^{-1}$. **11-K** is at 100 K also the most stable isomer compared to competing isomers based on our free-energy calculations. Thus, for $Au_{11}$ we assign the FIR-MPD spectra to the nonplanar capped-trigonal prism (**11-K**). A contribution of a minor fraction of the planar isomers **H+** (**11-H1** and **11-H2**) could explain the additional peaks seen in the FIR-MPD spectrum of $Au_{11}Kr$.

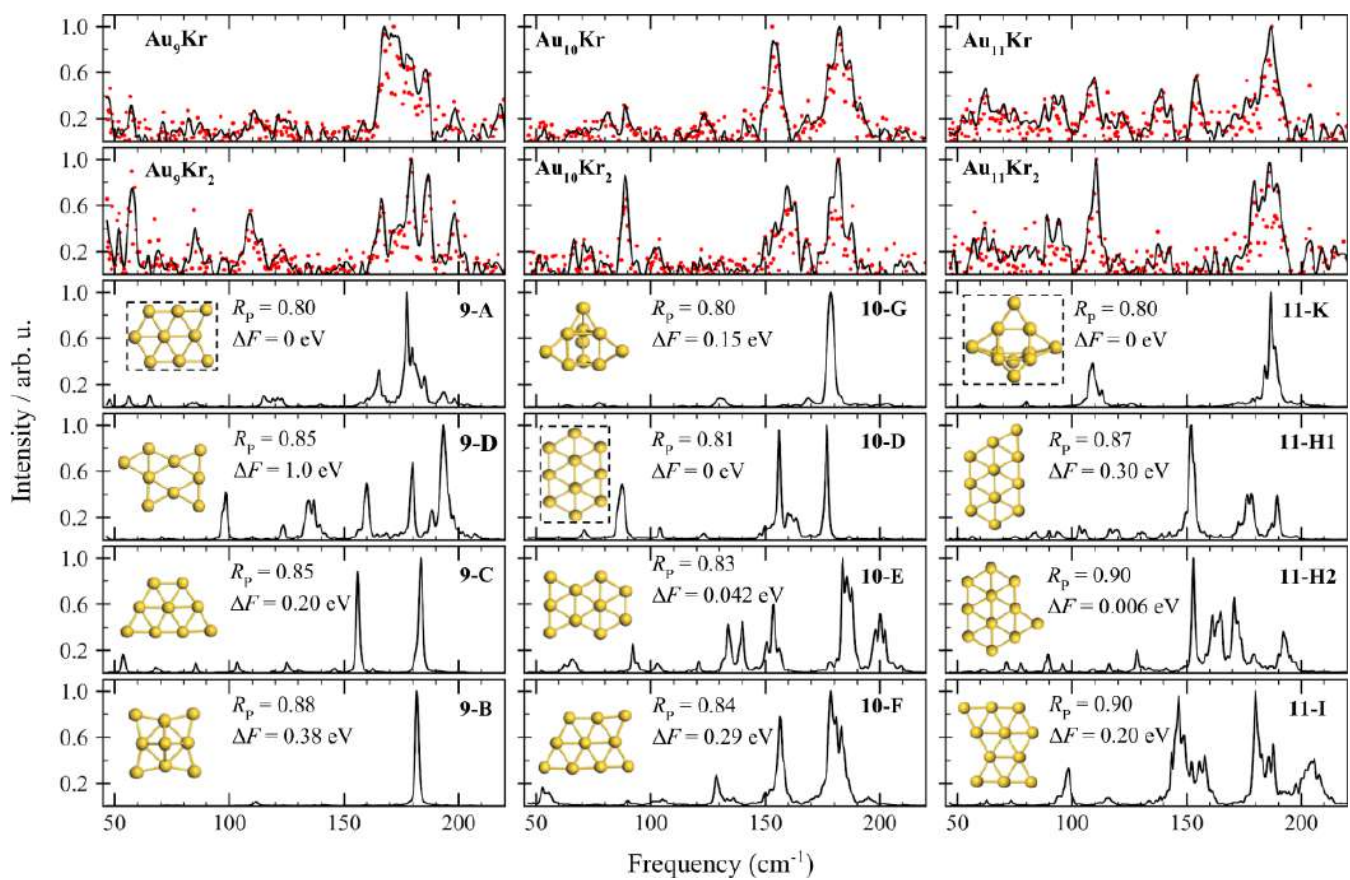

**FIG. 10.** Comparison of experimental FIR-MPD and theoretical anharmonic IR spectra for $Au_9$, $Au_{10}$, and $Au_{11}$ at 100 K. For each size, the two upper panels show the experimental spectra for the complexes with one and two Kr atoms and below theoretical IR spectra are shown. Pendry R-Factor ($R_P$) comparing theoretical IR spectra with experimental results from $Au_9Kr_2$, $Au_{10}Kr$, or $Au_{11}Kr_2$ are shown. Free-energy calculations ($\Delta F$, referenced to the lowest free-energy isomer) are shown at 100 K for each isomer. Structures surrounded by a black-dashed box are assigned to experimental FIR-MPD spectra based on $R_P$ and the free-energy calculations. The black lines in the experimental FIR-MPD spectra are smoothed spectra, and the red dots are the raw data.

**$Au_{12}$.** Our free-energy calculations suggest $Au_{12}$ most likely adopts a planar di-capped elongated-hexagon structure (**12-O**) at 100 K, whereas our theoretical IR spectra and experimental FIR-MPD spectra for $Au_{12}Kr$ and $Au_{12}Kr_2$ show good agreement for a nonplanar cup ($C_{2v}$) isomer (**FIG. 11, 12-P**). Although **12-V** has a similar IR spectrum to **12-P**, it is much less stable. This structural assignment is further supported by the theoretical harmonic IR spectra of **12-P**, which shows peaks around 135 cm$^{-1}$, 175 cm$^{-1}$, and 185 cm$^{-1}$ with similar intensity, whereas **12-V** does not (**FIG. S12** of the SI). **12-P** is found to be the second most stable isomer at 100 K (and the dominant one at 200 K and above). Interestingly, this nonplanar cup isomer **12-P** has the same structure as the prevailing isomer of the $Au_{12}^-$ anionic cluster [18,45]. One possibility is that **12-P** is kinetically trapped and cannot reach the thermodynamically accessible planar ground-state isomer (**12-O**) at 100 K [39]. The free-energy barrier to interconvert between nonplanar **12-P** and planar **12-O** is estimated to be appreciable even above 300 K, remaining greater than *ca.* 0.32 eV at 750 K. Another possibility is that inaccuracies in the predicted cluster isomer energy differences at zero Kelvin leads to a shift in the 2D/3D cross-over temperature of $Au_{12}$.

**$Au_{13}$.** For $Au_{13}$, previous DFT studies have suggested either the planar **13-R**, or one of the nonplanar **13-T** and **13-U** isomers is the most stable [44,57,67]. Our DFT calculations using PBE+MBD predict $Au_{13}$ to be most stable as the planar **13-R** at zero kelvin (and 100 K), whereas HSE06+MBD and RPA@PBE calculations predict the most stable structure is the nonplanar **13-T** at zero kelvin. At 100 K, we predict the nonplanar **13-T** is 29 meV uphill and **13-U** is 50 meV uphill from **13-R** using PBE+MBD. As evident in the IR spectra and molecular dynamics simulations, the nonplanar isomers **13-T** and **13-U** show significant dynamics even at 100 K. Since $Au_{11}$ and $Au_{12}$ were determined by FIR-MPD spectroscopy to adopt nonplanar structures, it is probable that $Au_{13}$ also prefers a nonplanar structure at 100 K. Thus, we tentatively assign the $Au_{13}$ structure in our FIR-MPD measurements to a mixture of the nonplanar **13-T** and **13-U** isomers. Although spectroscopic evidence is inconclusive on the precise structure of $Au_{13}$ at 100 K, it appears that $Au_{13}$ will adopt a variety of nonplanar structures.

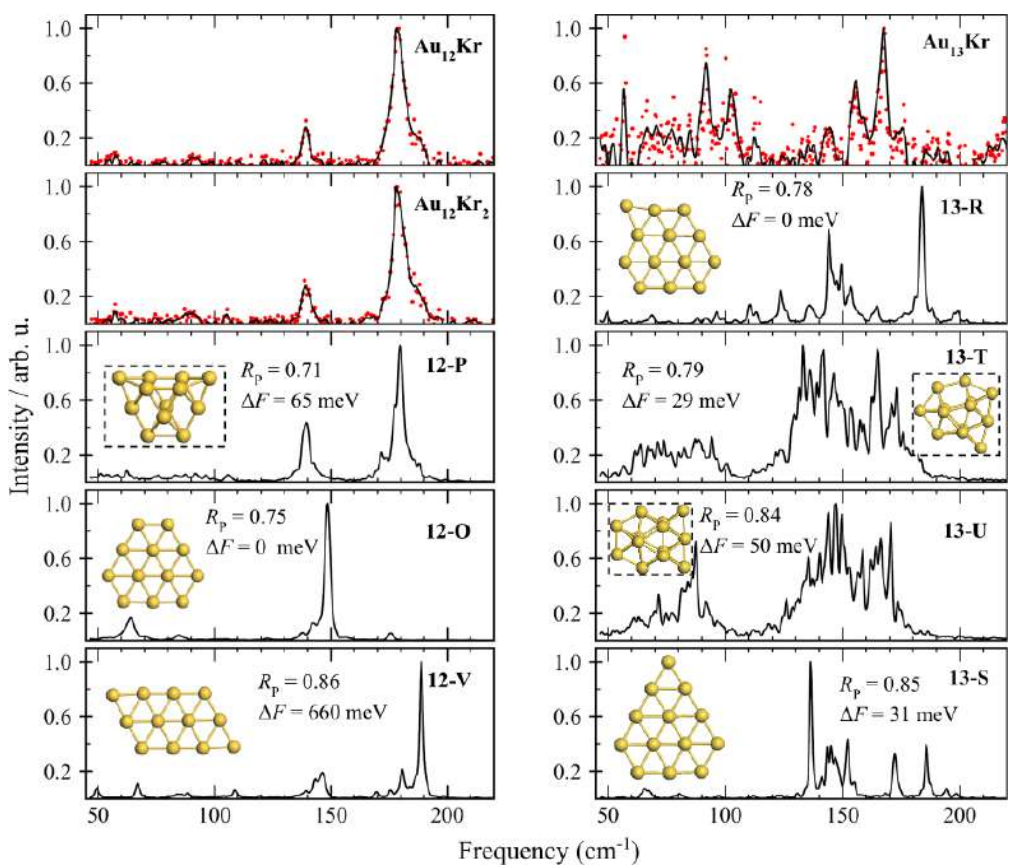

**FIG. 11.** Comparison of experimental FIR-MPD and theoretical anharmonic IR spectra for $Au_{12}$ and $Au_{13}$ at 100 K. For each size, the upper panel(s) show the experimental spectra for the complexes with Kr atoms and below theoretical IR spectra are shown. Pendry R-Factor ($R_P$) comparing theoretical IR spectra with experimental results from $Au_{12}Kr_2$ and $Au_{13}Kr$ are shown. Free-energy calculations ($\Delta F$, referenced to the lowest free-energy isomer) are shown at 100 K for each isomer. Structures surrounded by a black-dashed box are assigned to experimental FIR-MPD spectra based on $R_P$ and the free-energy calculations.

## SUMMARY

Accurate first-principles studies, properly considering both temperature effects and long-range van der Waals interactions, are used in conjunction with experimental far-IR multiple photon dissociation (FIR-MPD) spectroscopy to identify the structures of neutral gas-phase $Au_5$–$Au_{13}$ clusters at finite temperature. $Au_5$–$Au_8$ are predicted to adopt planar isomers at 100 K, which agrees with prior FIR-MPD measurements and first-principles calculations. Specifically, $Au_5$ is most stable as a trapezoid ($C_{2v}$),

Au$_6$ as a triangle (D$_{3h}$), Au$_7$ as an edge-capped triangle (C$_s$), and Au$_8$ as a 4-fold edge-capped square (D$_{4h}$). Our experimental FIR-MPD spectroscopy measurements and theoretical anharmonic IR spectra, combined with free-energy calculations, let us assign the structures of neutral Au$_9$-Au$_{13}$ clusters. Specifically, Au$_9$ (di-capped hexagon, C$_{2v}$) and Au$_{10}$ (elongated hexagon, D$_{2h}$) prefer planar isomers, whereas Au$_{11}$ (capped-trigonal prism, D$_{3h}$), Au$_{12}$ (cup, C$_{2v}$), and Au$_{13}$ (isomers **13-T** or **13-U**, C$_s$) adopt nonplanar isomers at 100 K. Thus, Au$_{10}$ can be considered the largest planar structure at 100 K. A Pendry R-factor analysis is used for the comparison of the theoretical and experimental spectra. Our study shows that for IR spectra it is often insufficient to assign theoretical IR spectra to experimental FIR-MPD spectra. Indeed, the Pendry R-Factor was originally developed for low energy electron diffraction analysis, thus it is also necessary to also consider the free energy of the gold clusters to aid structural assignments. We predict that increasing temperature generally stabilizes nonplanar gold clusters structures over planar structures due to a larger entropic contribution to the free energy, with the exception of Au$_{11}$ due to the large number of nearly energy-degenerate planar structures. Including long-range van der Waals interactions in DFT calculations has a pronounced effect on stabilizing nonplanar structures relative to planar structures and significantly improves ground-state energy predictions relative to calculations using the random-phase approximation. The impact of long-range van der Waals interactions on stabilizing nonplanar isomers over planar isomers is likely general for gas-phase clusters.

## Acknowledgements

BRG acknowledges support from the Alexander von Humboldt-Foundation and start-up funds from University of Michigan, Ann Arbor. The authors thank I.Y. Zhang, J. Hermann, V.V. Gobre, D.A. Goldsmith, and A. Tkatchenko for fruitful discussions.

# Table of Contents Graphic

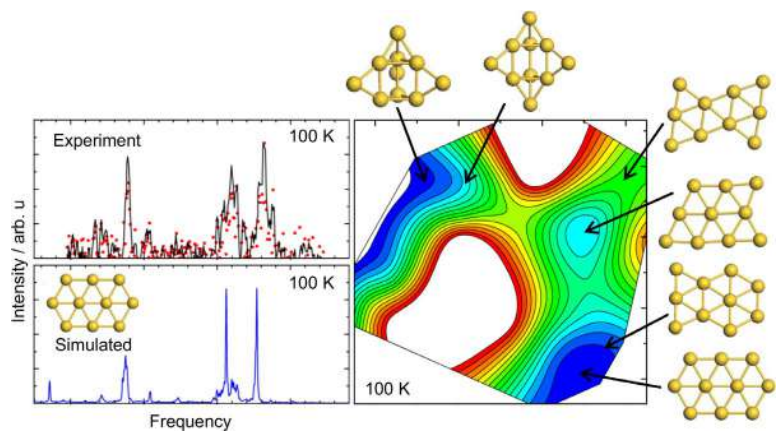

# Two-to-three dimensional transition in neutral gold clusters: the crucial role of van der Waals interactions and temperature

Bryan R. Goldsmith,[1,2*] Jacob Florian,[2] Jin-Xun Liu,[2] Philipp Gruene,[1] Jonathan T. Lyon,[1,3] David M. Rayner,[4] André Fielicke,[1*] Matthias Scheffler,[1] and Luca M. Ghiringhelli[1*]

[1]*Fritz-Haber-Institut der Max-Planck-Gesellschaft, Faradayweg 4-6, D-14195 Berlin, Germany*
[2]*Department of Chemical Engineering, University of Michigan, Ann Arbor, MI 48109-2136, USA*
[3]*Department of Chemistry and Biochemistry, Kennesaw State University, 370 Paulding Avenue NW, MD#1203, Kennesaw, GA 30144, USA*
[4]*National Research Council, 100 Sussex Drive, Ottawa, Ontario, Canada K1A OR6*
Corresponding authors: bgoldsm@umich.edu, fielicke@fhi-berlin.mpg.de, ghiringhelli@fhi-berlin.mpg.de

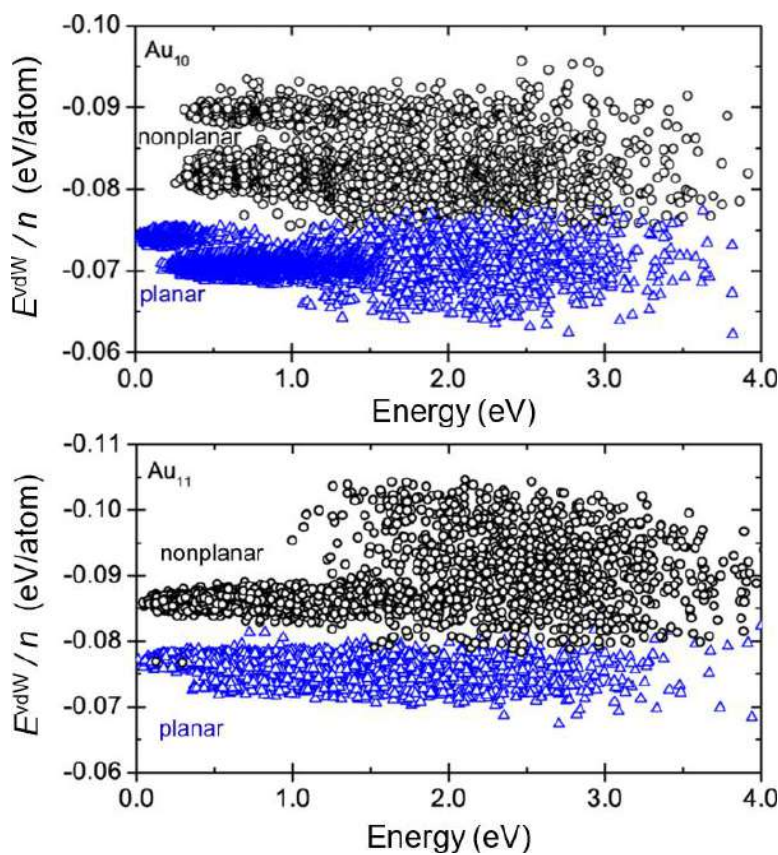

**FIG. S1.** Many-body dispersion energy per atom ($E^{vdW}/n$) vs. gold cluster total energy obtained from replica-exchange *ab initio* molecular dynamics for $Au_{10}$ (top) and $Au_{11}$ (bottom). Planar and nonplanar cluster isomers are distinguished based on their radius of gyration (nonplanar clusters have a smaller radius of gyration than planar structures) and designated as blue triangles and black circles, respectively. The MBD energy was computed using PBE+MBD as described in the main text.

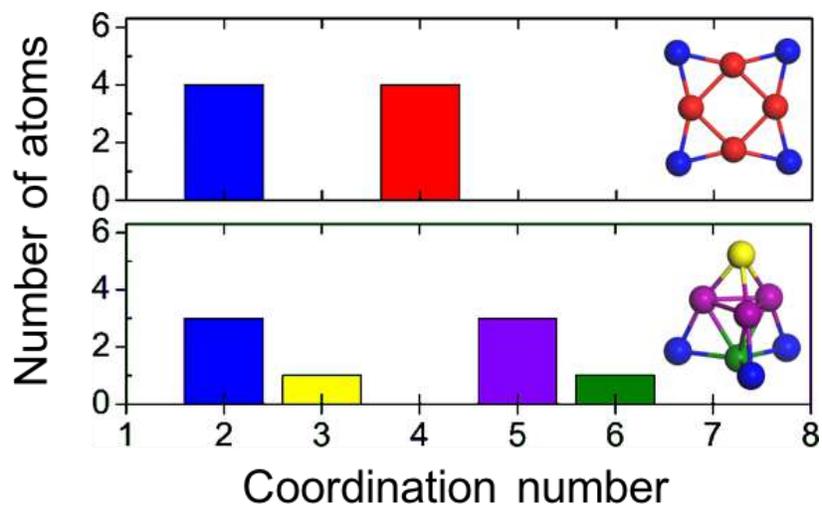

**FIG. S2.** Bond coordination histograms of the most stable planar and nonplanar $Au_8$ isomers. The bond coordination histogram serves as a unique structural fingerprint because it represents each gold isomer as having a certain number of atoms with some bond coordination number. The coordination similarity metric is obtained by constructing a bond coordination histogram for each cluster configuration and computing the cosine distance between each bond coordination histogram (represented as a vector) with respect to the ground state reference. The cosine distance is chosen instead of the Euclidean distance because it is less sensitive to noise (i.e., small perturbations in structure). The gold cluster equilibrium bond distance was set to 2.72 Å for defining the coordination histogram. Color legend: The colors of the gold cluster atoms correspond to their coordination number.

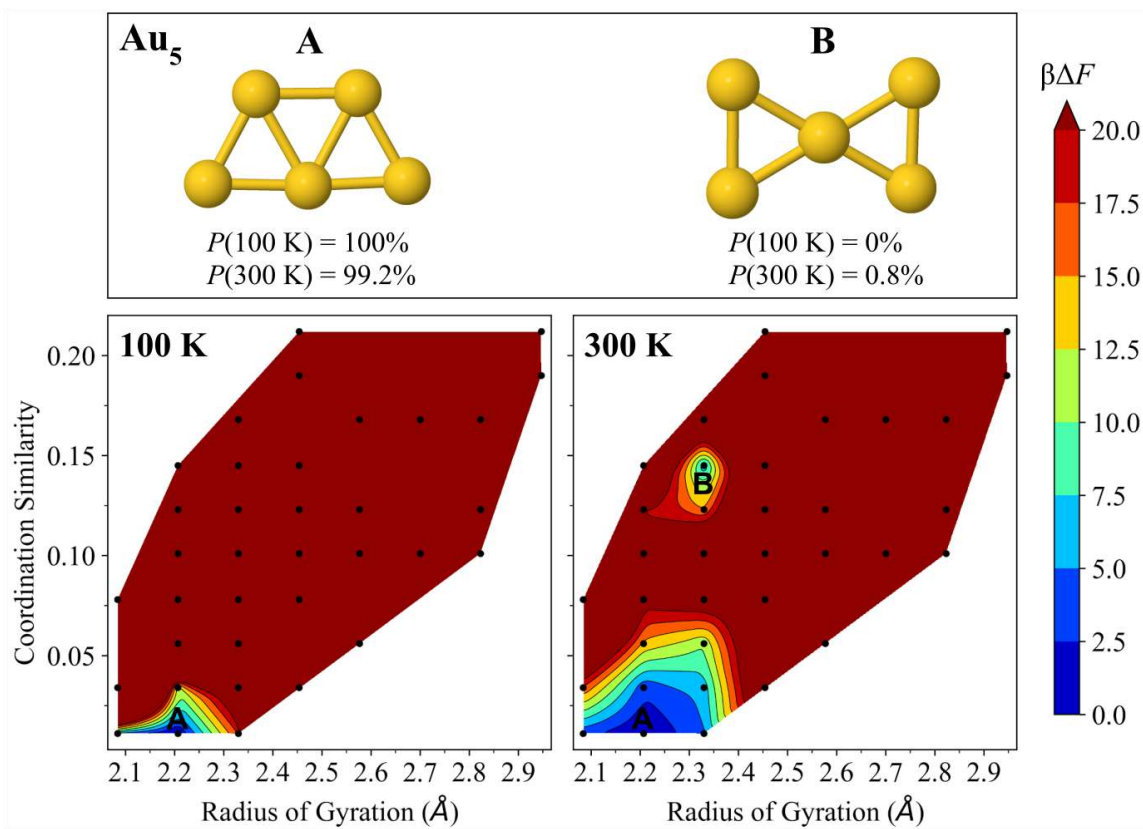

**FIG. S3.** Free-energy surfaces of Au$_5$ at 100 K (left) and 300 K (right). The dominant isomer **5-A** (trapezoid, C$_{2v}$) and competing isomer **5-B** (bow-tie, D$_{2h}$) have their Boltzmann probabilities $P(T)$ specified at each temperature, as well as their locations marked on the free-energy surfaces. The color bar denotes dimensionless free-energy values ($\beta\Delta F$), where $\beta$ is $1/k_B T$. Dark red regions have $\beta\Delta F \geq 20$. The black points indicate sampled cluster geometries having the binned (coordination similarity, radius of gyration) value.

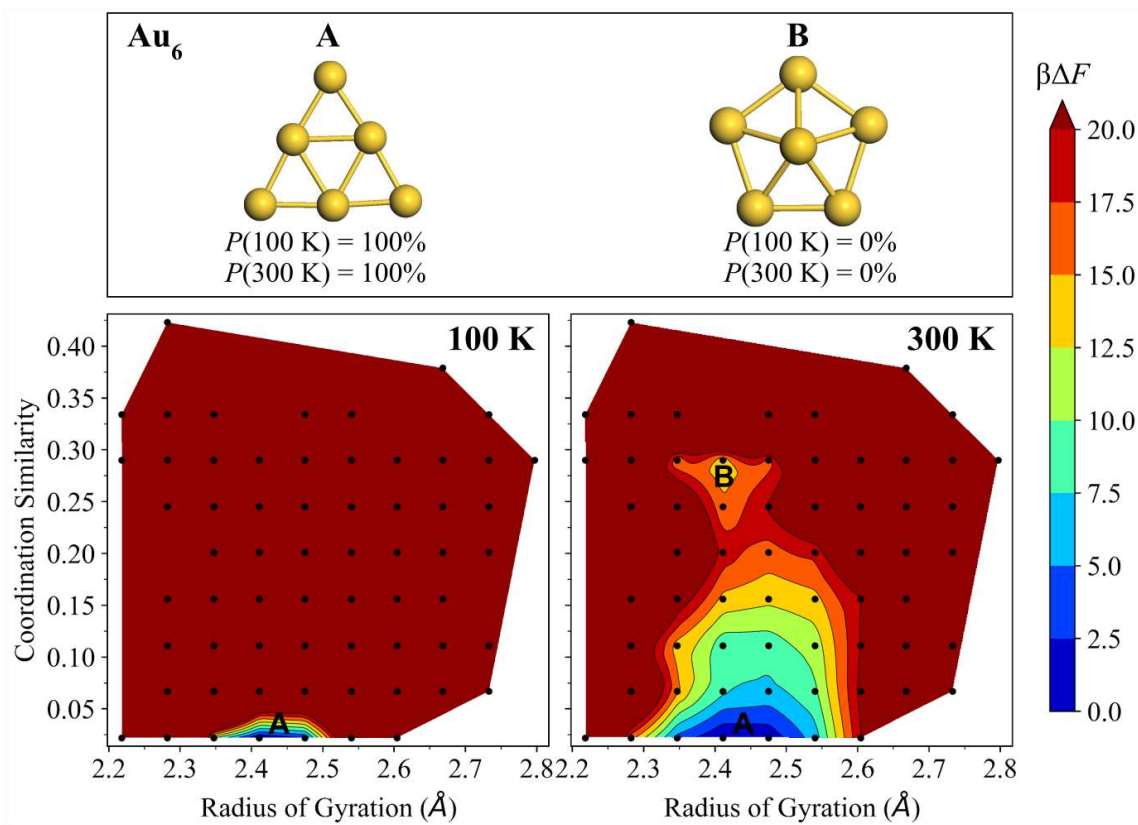

**FIG. S4.** Free-energy surfaces of Au$_6$ at 100 K (left) and 300 K (right). The dominant isomer **6-A** (triangle, D$_{3h}$) and second most stable isomer **6-B** (capped pentagon, C$_{5v}$) have their Boltzmann probabilities $P(T)$ specified at each temperature, as well as their locations marked on the free-energy surfaces. The color bar denotes dimensionless free-energy values ($\beta\Delta F$), where $\beta$ is $1/k_BT$. Dark red regions have $\beta\Delta F \geq 20$. The black points indicate sampled cluster geometries having the binned (coordination similarity, radius of gyration) value.

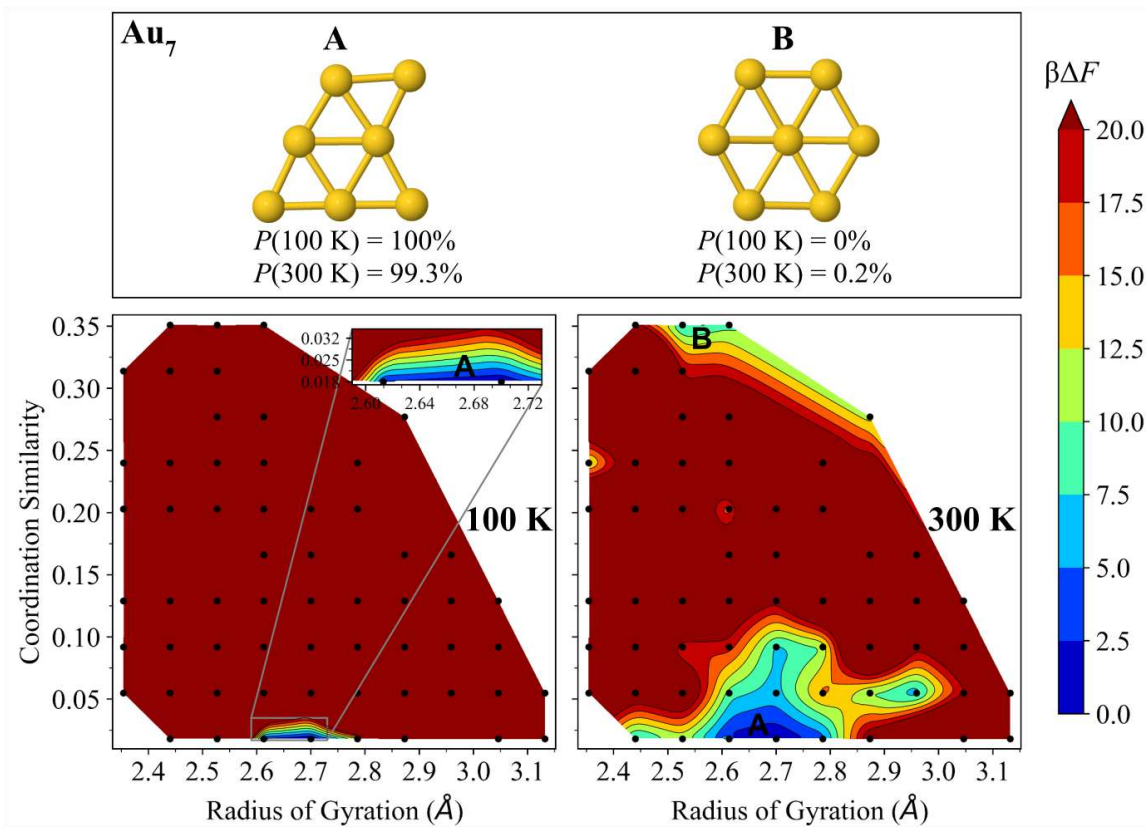

**FIG. S5.** Free-energy surfaces of Au$_7$ at 100 K (left) and 300 K (right). The dominant isomer **7-A** (edge-capped triangle, C$_s$) and competing isomer **7-B** (hexagon, D$_{6h}$) have their Boltzmann probabilities $P(T)$ specified at each temperature, as well as their locations marked on the free-energy surfaces.

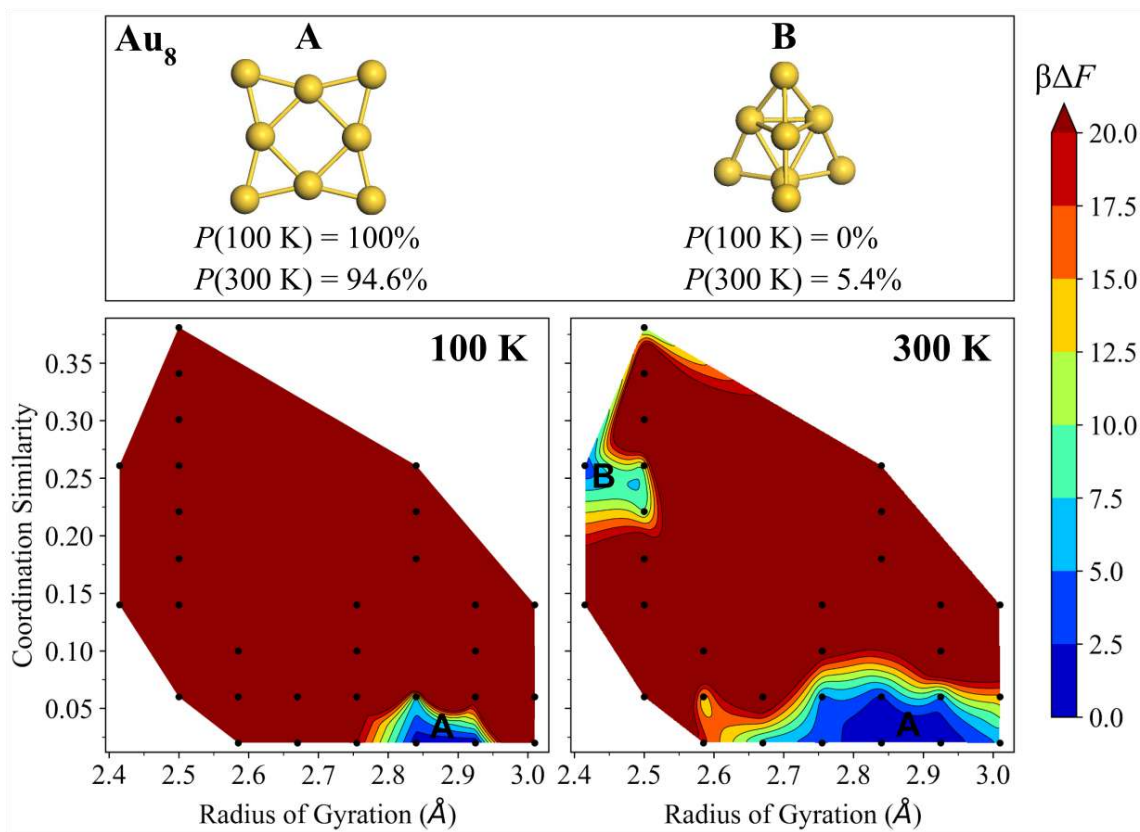

**FIG. S6.** Free-energy surfaces of Au$_8$ at 100 K (left) and 300 K (right). The dominant isomer **8-A** (4-fold edge-capped square, D$_{4h}$) and competing isomer **8-B** (triangular pyramid, C$_{3v}$) have their Boltzmann probabilities $P(T)$ specified at each temperature, as well as their locations marked on the free-energy surfaces.

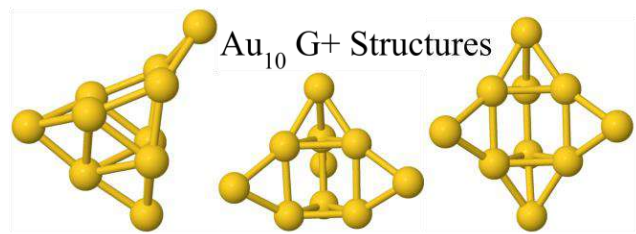

**FIG. S7.** Metastable Au$_{10}$ structures corresponding to **10-G+** on the Au$_{10}$ free-energy surface (FIG. 6 in the main text and FIG. S8 of the SI).

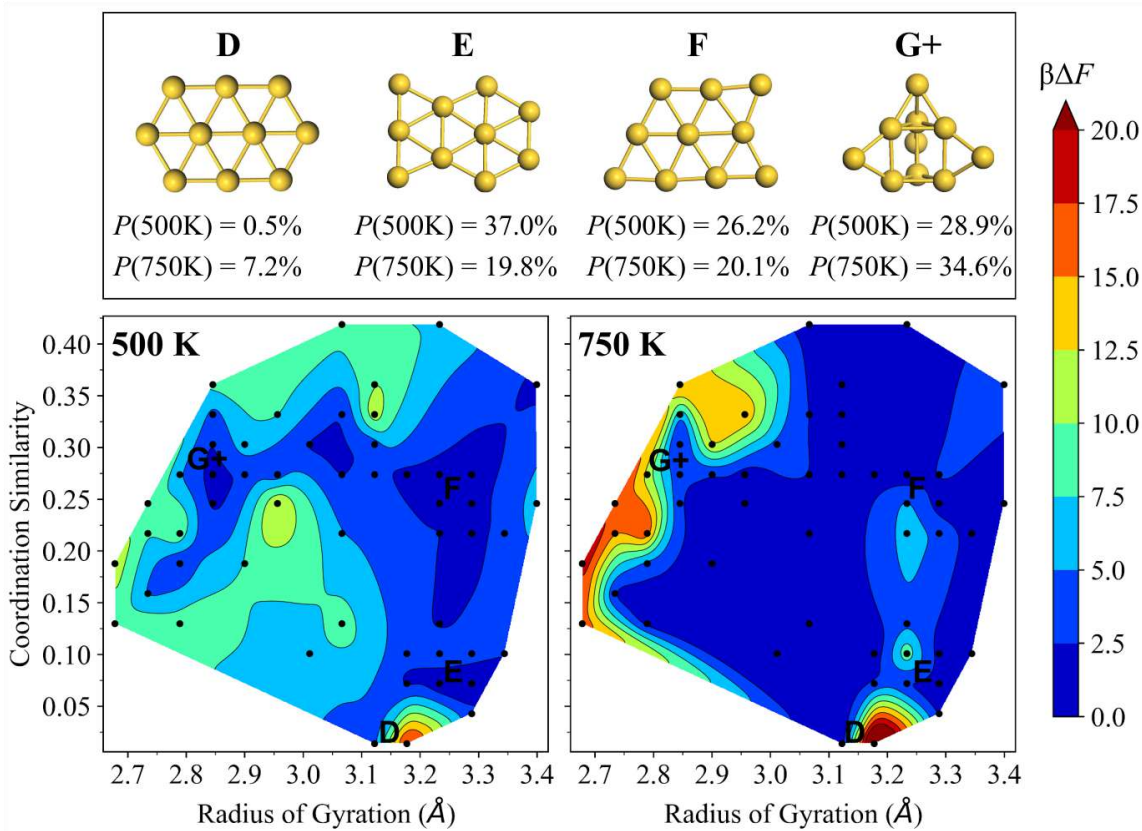

**FIG. S8.** Free-energy surfaces of $Au_{10}$ at 500 K (left) and 750 K (right). The dominant isomers **10-D** (elongated hexagon, $D_{2h}$), **10-E** (tri-capped hexagon, $C_{2h}$), **10-F** (capped trapezoid, $C_s$), and **10-G+** have their Boltzmann probabilities $P(T)$ specified at each temperature, as well as their locations marked on the free-energy surfaces. **10-G+** refers to multiple nonplanar isomers close to each other on the free-energy surface (FIG. S7 of the SI). The color bar represents dimensionless free-energy values ($\beta\Delta F$), where $\beta$ is $1/k_B T$. Dark red regions have $\beta\Delta F \geq 20$. The black points indicate sampled cluster geometries having the binned (coordination similarity, radius of gyration) value. Isomers with $P(300\ K) < 1\%$ are not shown.

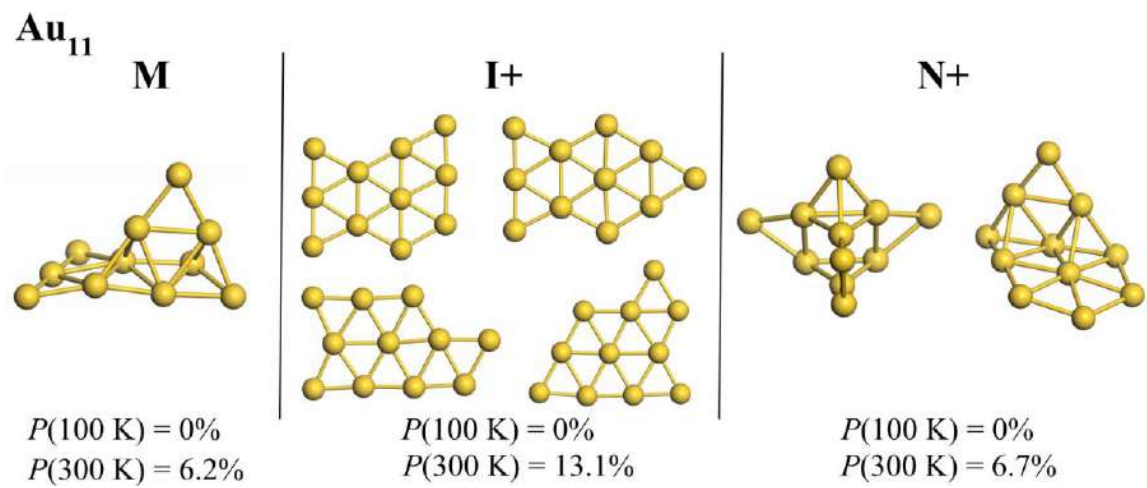

**FIG. S9.** Metastable Au$_{11}$ structures corresponding to **11-I+, 11-M, 11-N+** on the Au$_{11}$ free-energy surface (FIG. 7 in the main text).

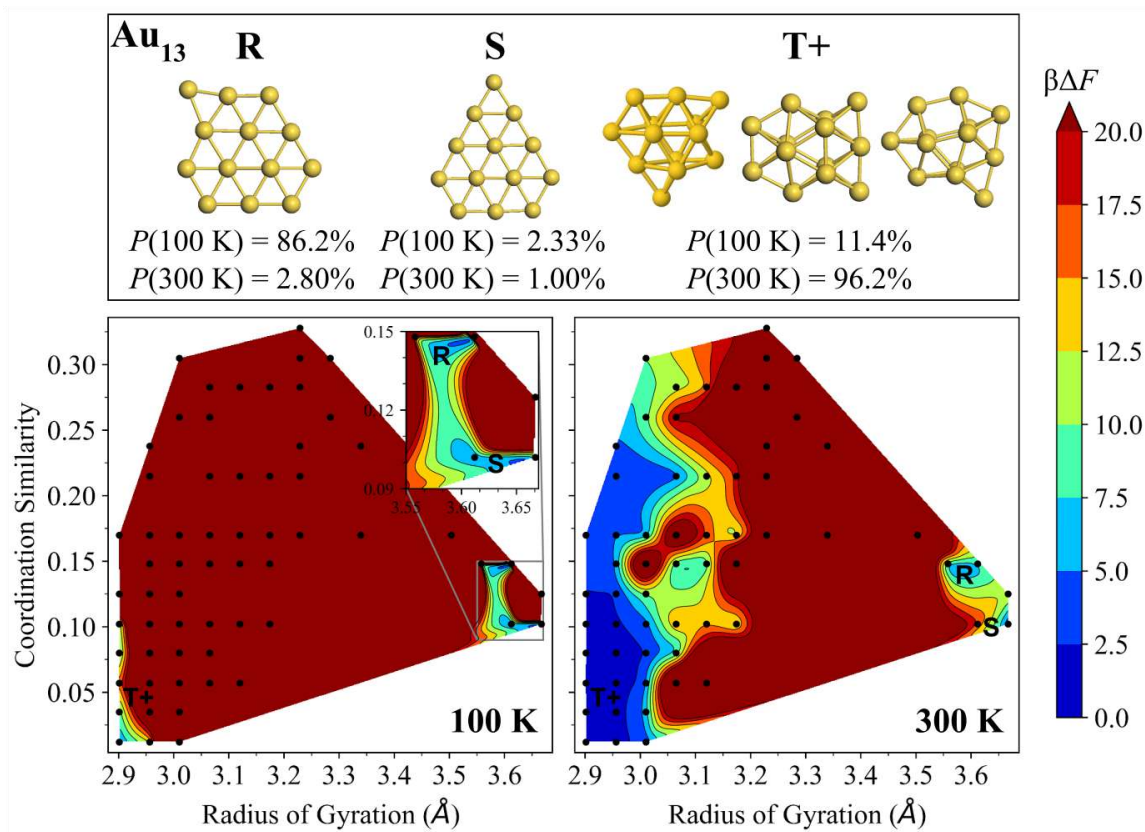

**FIG. S10.** Free-energy surfaces of Au$_{13}$ at 100 K (left) and 300 K (right). The dominant isomers **13-R**, **13-S**, and **13-T+** have their Boltzmann probabilities $P(T)$ specified at each temperature, as well as their locations marked on the free-energy surfaces. The color bar denotes dimensionless free-energy values ($\beta\Delta F$), where $\beta$ is $1/k_B T$. Au$_{13}$ structures (**13-T+**) are not shown for clarity, but match those of E. C. Beret, L. M. Ghiringhelli, and M. Scheffler, *Farad. Discuss.* **152**, 153 (2011).

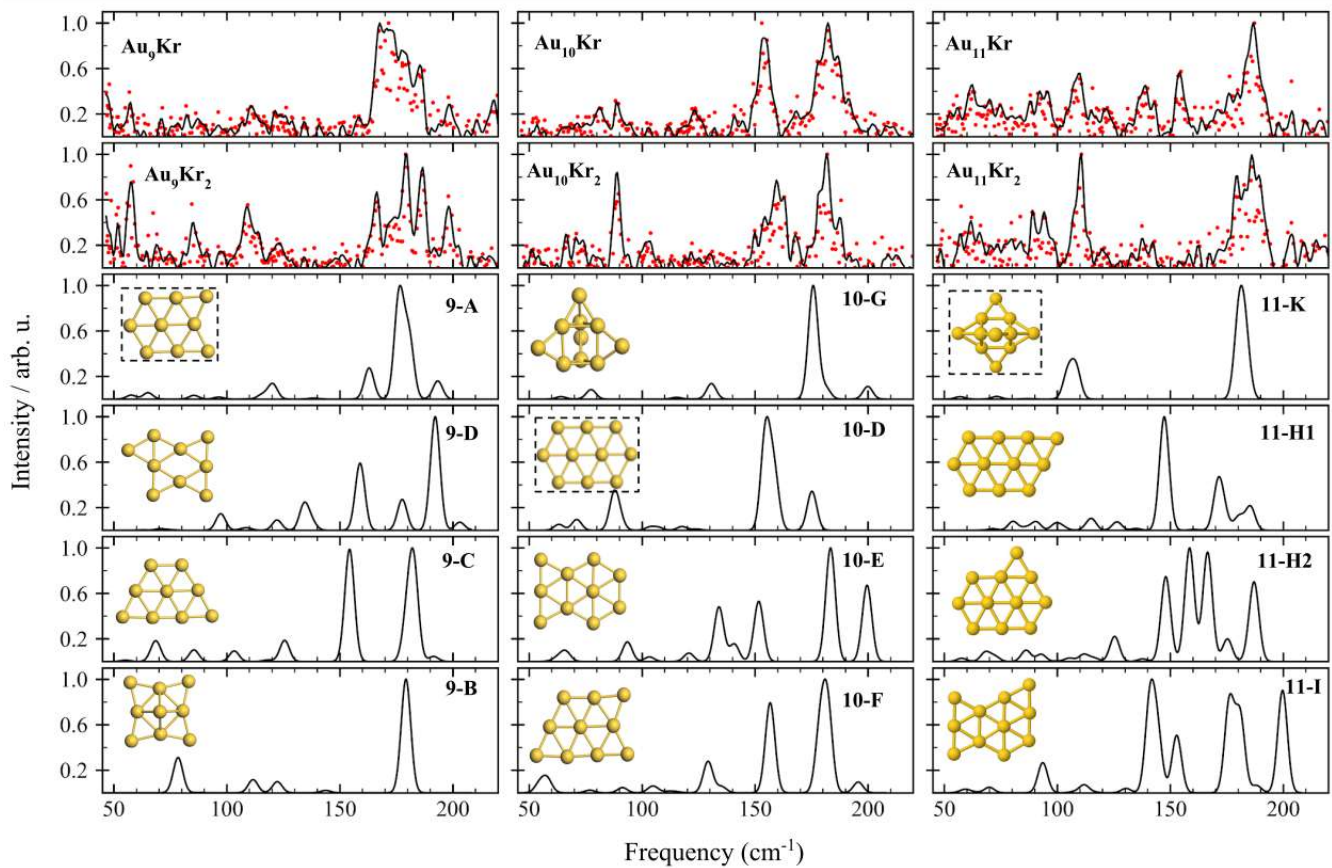

**FIG. S11.** Comparison of experimental FIR-MPD and theoretical harmonic IR spectra (at 0 kelvin) for $Au_9$, $Au_{10}$, and $Au_{11}$. For each size, the two upper panels show the experimental spectra for the complexes with a single and two Kr atoms and below the theoretical IR spectra are shown. A rigid shift of 8 cm$^{-1}$ is used for $Au_9$ and $Au_{10}$, whereas a rigid shift of 12 cm$^{-1}$ is used for $Au_{11}$. Gaussian broadening with a 5 cm$^{-1}$ FWHM is used. Structures surrounded by a black-dashed box are assigned to experimental FIR-MPD spectra based on Pendry R-factor analysis and the free-energy calculations.

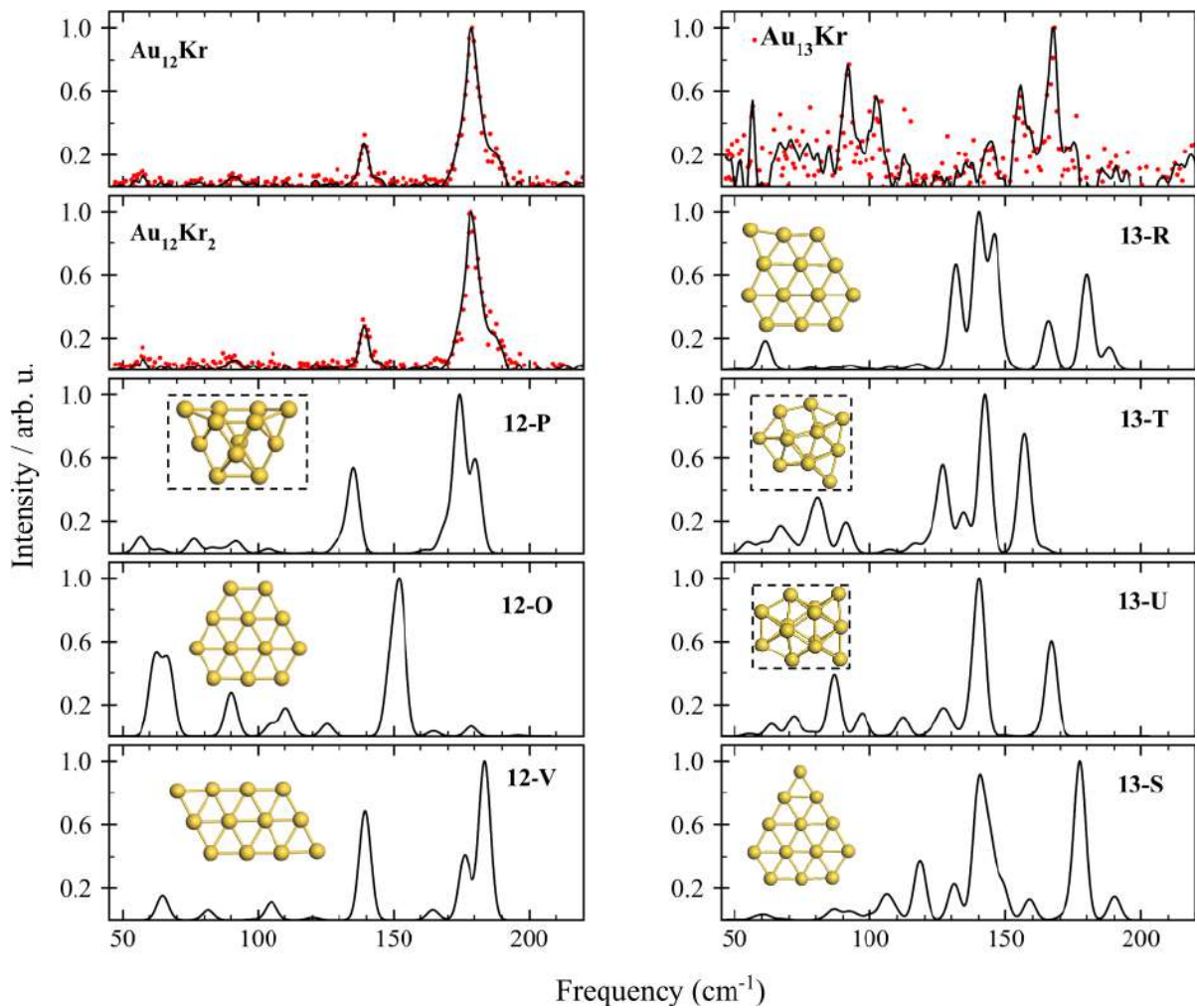

**FIG. S12.** Comparison of experimental FIR-MPD and theoretical harmonic IR spectra (at 0 kelvin) for $Au_{12}$ and $Au_{13}$. For each size, the top panels show the experimental spectra for the complexes with one or two Kr atoms and below the theoretical IR spectra are shown. A rigid shift of 3 cm$^{-1}$ and 8 cm$^{-1}$ is used for $Au_{12}$ and $Au_{13}$ simulated spectra, respectively. Gaussian broadening with a 5 cm$^{-1}$ FWHM is used. Structures surrounded by a black-dashed box are assigned to experimental FIR-MPD spectra based on Pendry R-factor analysis and the free-energy calculations.

**Table S1.** Energy difference (eV) between the lowest energy planar (2D) and nonplanar (3D) isomers for $Au_5$–$Au_{13}$ using various density functional approximations and beyond. The values in this table correspond to FIG. 2 in the main text. The planar and nonplanar gold clusters used to obtain the energy differences correspond to those shown in FIG. 1 of the main text. The most stable planar and nonplanar 2D and 3D isomers at 0 K did not change for the considered DFAs.

| Cluster size | PBE | PBE+TS | PBE+MBD | HSE06 | HSE06+TS | HSE06+MBD | RPA@PBE |
|---|---|---|---|---|---|---|---|
| $Au_5$ | −0.964 | −0.981 | −0.929 | −0.973 | −0.992 | −0.946 | −0.901 |
| $Au_6$ | −0.980 | −0.949 | −0.938 | −0.937 | −0.912 | −0.903 | −0.957 |
| $Au_7$ | −0.296 | −0.287 | −0.267 | −0.246 | −0.237 | −0.220 | −0.199 |
| $Au_8$ | −0.526 | −0.464 | −0.416 | −0.522 | −0.458 | −0.427 | −0.445 |
| $Au_9$ | −0.265 | −0.187 | −0.218 | −0.234 | −0.168 | −0.189 | −0.154 |
| $Au_{10}$ | −0.249 | −0.136 | −0.179 | −0.221 | −0.135 | −0.154 | −0.118 |
| $Au_{11}$ | −0.096 | 0.015 | 0.001 | 0.006 | 0.090 | 0.098 | 0.126 |
| $Au_{12}$ | −0.525 | −0.348 | −0.231 | −0.394 | −0.251 | −0.135 | −0.108 |
| $Au_{13}$ | −0.451 | −0.156 | −0.034 | −0.278 | −0.025 | 0.092 | 0.089 |
| MAE | 0.187 | 0.093 | 0.071 | 0.130 | 0.057 | 0.030 | -- |
| STDev | 0.166 | 0.086 | 0.041 | 0.110 | 0.045 | 0.014 | -- |

*Mean absolute error (MAE) and standard deviation (STDev) with respect to RPA@PBE.

**Negative values indicate that a planar structure is preferred over the nonplanar structure.